\documentclass[a4paper]{aa}

\usepackage{graphicx}
\usepackage{txfonts,bm}

\newcommand{\dd}{\mathrm{d}}

\newcommand{\mat}{\mathrm{m}}
\newcommand{\rad}{\mathrm{r}}
\newcommand{\de}{\mathrm{de}}
\newcommand{\q}{\mathrm{Q}}
\newcommand{\ap}{\mathrm{ap}}
\newcommand{\pivot}{\mathrm{pivot}}
\newcommand{\etal}{et al. }

\begin{document}

 \title{Tracking quintessence by cosmic shear}

   \subtitle{Constraints from VIRMOS-Descart and CFHTLS\thanks{Based
   on observations obtained with MegaPrime/MegaCam, a joint project
   of CFHT and CEA/DAPNIA, at the Canada-France-Hawaii Telescope (CFHT)
   which is operated by the National Research Council (NRC) of Canada,
   the Institut National des Sciences de l'Univers of the Centre National
   de la Recherche Scientifique (CNRS) of France, and the University of
   Hawaii. This work is based in part on data products produced at
   TERAPIX and the Canadian Astronomy Data Centre as part of the
   Canada-France-Hawaii Telescope Legacy Survey, a collaborative
   project of NRC and CNRS.} and future prospects}

 \author{Carlo Schimd\inst{1,2}
       \and
           Ismael Tereno\inst{2,3}
       \and
           Jean-Philippe Uzan\inst{2}
       \and
           Yannick Mellier\inst{2,4}
       \and \\
           Ludovic van Waerbeke\inst{5}
       \and
           Elisabetta Semboloni\inst{2}
       \and
           Henk Hoekstra\inst{6}
       \and
           Liping Fu\inst{2}
       \and
           Alain Riazuelo\inst{2}
          }

   \offprints{{\tt carlo.schimd@cea.fr}}

   \institute{DSM/DAPNIA, CEA/Saclay,
              91191 Gif sur Yvette cedex, France
         \and
              Institut d'Astrophysique de Paris, UMR7095 CNRS,
              Universit\'e Pierre~\&~Marie Curie - Paris,
              98 bis bd Arago, 75014 Paris, France
         \and
              Departamento de Fisica, Universidade de Lisboa,
              1749-016 Lisboa, Portugal
         \and
              Observatoire de Paris - LERMA,
              61 avenue de l'Observatoire, 75014 Paris, France
         \and
              Department of Physics and Astronomy, University of British Columbia,
              6224 Agricultural Road, Vancouver V6T 1Z1, Canada
         \and
              Department of Physics and Astronomy, University of Victoria,
              Victoria V8P 52C, Canada
              }

   \date{Received \today; accepted ...}


\abstract
{Dark energy can be investigated in two complementary ways, by considering either general
parameterizations or physically well-defined models. This article follows the second route and
explores the observational constraints on quintessence models where the acceleration of our
universe is driven by a slow-rolling scalar field. The analysis focuses on cosmic shear, combined
with type~Ia supernovae data and cosmic microwave background observations.}
{This article examines how weak lensing surveys can constrain dark energy, how they complement
supernovae data to lift some degeneracies and addresses some issues regarding the limitations due
to the lack of knowledge concerning the non-linear regime.}
{Using a Boltzmann code that includes quintessence models and the computation of weak lensing
observables, we determine the shear power spectrum and several two-point statistics. The
non-linear regime is described by two different mappings. The likelihood analysis is based on a
grid method. The data include the ``gold set'' of supernovae Ia, the WMAP-1 year data and the
VIRMOS-Descart and CFHTLS-deep and -wide data for weak lensing. This is the first analysis of
high-energy motivated dark energy models that uses weak lensing data. We explore larger angular
scales, using a synthetic realization of the complete CFHTLS-wide survey as well as next
space-based missions surveys.}
{Two classes of cosmological parameters are discussed: $i)$ those accounting for quintessence
affect mainly geometrical factors; $ii)$ cosmological parameters specifying the primordial
universe strongly depend on the description of the non-linear regime. This dependence is addressed
using wide surveys, by discarding the smaller angular scales to reduce the dependence on the
non-linear regime. Special care is payed to the comparison of these physical models with
parameterizations of the equation of state. For a flat universe and a quintessence inverse power
law potential with slope $\alpha$, we obtain $\alpha<1$ and $\Omega_{\q0}=0.75^{+0.03}_{-0.04}$
at 95\% confidence level, whereas $\alpha=2^{+18}_{-2}$, $\Omega_{\q0}= 0.74^{+0.03}_{-0.05}$
when including supergravity corrections.}
{}

   \keywords{Gravitational lensing. Cosmology: theory -- cosmological parameters. Methods --
    data analysis.}
   \maketitle

\section{Introduction}\label{sec1}

Cosmological observations provide increasing compelling evidences that the expansion of the
universe is accelerating and that the cosmic history of the universe seems today dominated by
another component than its matter and radiation content (see e.g.~\cite{peebles03};
\cite{caroll01}; \cite{quint3}; \cite{pubook}, chap.~12 for reviews and references therein). If
so, one of the most challenging issue of fundamental physics is to understand the cause of this
acceleration, a question often referred to as the nature of the {\it dark energy}.
 Various solutions, from the introduction of a new type of matter to a modification of general
relativity to describe the gravitation interaction, have been considered. A classification of
these models with some relevant observational tests that can help to distinguish between each
class from their underlying new physics is  discussed in Uzan (2004) and Uzan, Aghanim \& Mellier
(2004).

Dark energy appears in the Friedmann equations through its {\it effective} density and pressure.
Data are usually interpreted assuming the validity of the Copernician principle (so that the
dynamics of spacetime is completely described by a single function, the scale factor $a$) and the
validity of Einstein equations (and thus the standard Friedmann equations), so that the density
and pressure of the dark energy component are defined by $\rho_\de = (3/8\pi
G)(H^2+K/a^2)-\rho_\mat-\rho_\rad$ and $P_\de=(-1/8\pi G)(\ddot a/a + H^2 +K/a^2)$, where $H=\dot
a/a$ is the Hubble parameter and a dot refers to a derivative with respect to the cosmic time,
while $\rho_\mat$ and $\rho_\rad$ are the density of pressureless matter and radiation,
respectively. It follows that the equation of state of the dark energy corresponds to
\begin{equation}
 3\Omega_\de w=-1+\Omega_K + 2q,
\end{equation}
where $q=-a\ddot a/\dot a^2$ is the deceleration parameter and $\Omega_K=-K/a^2H^2$. From this
point of view, $w$ characterizes the dynamics of the cosmic expansion. More precisely, it
parameterizes the deviation, $H(z)-\bar H(z)$, between the Hubble function of the observed
Universe, $H(z)$, and that predicted for a universe filled only with pressureless matter and
radiation, $\bar H(z)$. It is therefore equivalent to specify $w(z)$ or $H(z)-\bar H(z)$. However,
when general relativity is assumed to describe gravity, $w$ reduces to $P_\de/\rho_\de$ so that,
in addition to the deviation from $\bar H(z)$, it also gives some insight on the properties of
dark energy (see e.g.~Martin, Schimd \& Uzan (2005) for a case in which $w$ does not reduce t
the equation of state of a matter component).

Although effective equation of state derived from observations is a key empirical
information on the rough nature of dark energy,  a detailed description of its properties
demands more thoughtful data interpretation. For example, all geometrical observables rely
on the integration of the Hubble parameter, hence on a double integration of the equation of
state $w$, that eventually dilutes or totally washes out its possible redshift dependence.
If $w$ is close to $-1$, as observations tend to indicate, then it is in general difficult
to demonstrate by geometrical tests that $w\not=-1$ or that $\dd w/\dd z\not=0$; both would
exclude a pure cosmological constant. Exploring early properties of dark energy
models would be even more challenging since for $w \simeq -1$ the ratio between the matter
and dark energy densities scales approximatively as $(1+z)^3$ so that dark energy is
dynamically negligible at redshift $z\gtrsim 2$. It leaves little freedom to determine the
scaling of the dark energy density and to demonstrate that it is not properly described by a
power law [$(1+z)^n$], as would be the case for a constant $w$ (see e.g.~\cite{kujat}).

From the theoretical point of view two routes can be followed. One can either exhibit a
general ``model-independent'' parameterization of the equation of state of the dark energy,
as discussed in the previous paragraph, or rely on a completely specified theoretical
models. A useful parameterization has to be realistic, in the sense that it should reproduce
predictions of a large class of models, it has to minimize the number of free parameters and
to be simply  related to the underlying physics (see e.g.~Linder \& Huterer 2005). Because the result of
the analysis will necessarily have some amount of parameterization dependence (\cite{bck}),
choosing the specified physical model strategy seems preferable to break degeneracies. In
particular, it enables to compute without any ambiguity their signature both in low and high
redshift surveys, such as the cosmic microwave background (CMB). The increasingly
flourishing number of models hampers to provide a comprehensive set of unambiguous
predictions to constrain physical models one by one with present-day observations, but there
are still several benefits in exploring dark energy this way, in particular when  weak
lensing surveys are used together with CMB observations. This is deeply related to the
evolution of dark energy properties and the growth rate of structure with look-back time, as
discussed below.

At low redshift, $w$ suffices to get observable that are all functions of $H(z)$ (see for
example~\cite{peebles93}; \cite{pubook}). This is the case of all background quantities
(e.g. luminosity distance, angular distance, look-back time, etc.) as well as of the linear growth
factor of density perturbations. It follows that the equation of state encodes all relevant
information, provided the amplitude of the power spectrum is calibrated by adding a new
parameter, $\sigma_8$, the variance of the density perturbation on a scale of
$8\,h^{-1}$~Mpc (see e.g.~\cite{berben}; \cite{ludoben}; \cite{doran}). However, as far as
weak lensing is concerned, it was shown (\cite{ludoben}; \cite{berben}) that, for a fixed
redshift of the sources, the modification of the growth factor in the linear regime was
degenerate with the normalization factor. Hence normalizing on the CMB avoids this problem,
and at the same time it is important to describe the non-linear regime. The use of CMB
together with weak lensing data is therefore a logical way to constrain specified
theoretical models, beyond the description by an empirical equation of state.

At higher redshift, and in particular to relate the amplitude of the matter density power spectrum
to the one of the primordial power spectrum, one would need to include a description of the
evolution of the perturbations of dark energy. In particular, this effect becomes increasingly
important as $w$ approaches zero (\cite{berben}). Note that dark energy perturbations have a
non-adiabatic component that also requires a detailed model to be described. This depends on the
physical model of dark energy and cannot be incorporated in a simple model-independent way.

While the ability of lensing data to constrain the equation of state of dark energy has been
widely studied (\cite{ludoben}; \cite{hujain}; \cite{jtay}), there have been very few
analysis with real data. Hoekstra \etal (2005) and Semboloni \etal (2005) used the CFHTLS wide
and deep data to constrain a constant equation of state. Jarvis \etal (2005) analyzed the 75~square
degrees CTIO lensing survey, combined with type~Ia supernovae (Sn~Ia) data and CMB, assuming
a constant equation of state and a parameterization of the form proposed by~\cite{para3} and
then by~\cite{para3bis}. In this article, we consider a class of completely defined
quintessence models, realized by a self-interacting scalar field. Hence all observational
signatures (Sn~Ia, lensing, CMB) can be explicitly computed, with no ambiguity in the way to
deal with the perturbations of dark energy. This theoretical extension of the standard
$\Lambda$CDM model involves only one additional parameter, needed to characterize the
self-interacting potential of the quintessence field. It follows that, as discussed above,
we will be able to normalize our initial power spectrum on the CMB angular power spectrum
and, as a consequence, $\sigma_8$ and any possible dependence on the shape of the analytical
fit of the transfer function will disappear from our discussion; the value of $\sigma_8$
will be an output of each models. The problem of the pivot redshift (see \S~\ref{subsec22})
that appears when combining different datasets also disappear in that approach. As a
conclusion, this approach is very efficient in terms of the number of extra-parameters and
of the interpretation of the data. Let us emphasize that, even though we also consider Sn~Ia
and CMB data, we will focus on weak lensing - cosmic shear data. This article, being the
first analysis of the CFHTLS data for dark energy studies, illustrates the power and the
problems of lensing survey in studying dark energy.

The article is organized as follows. In Section~\ref{sec2}, we define the quintessence
models we are considering and we recall their main properties. We also compare them to
various parameterizations proposed in the literature. Section~\ref{sec3} focuses on cosmic
shear. After a reminder on theoretical issues, we describe the weak lensing data used for
our analysis and the computational pipeline, finally we outline the likelihood analysis on
real (\S~\ref{sec3d}) and synthetic (\S~\ref{sec3e}) datasets. In Section~\ref{sec4} we
combine weak lensing data with type Ia supernovae and CMB temperature anisotropies;
Figure~\ref{fig:4a} and Table~\ref{tab:4} summarize the constraints on quintessence
parameters. We finish, in Section~\ref{sec5}, by an estimation of the proficiency of two
possible space-based wide field imagers to unveil the nature of dark energy.


\section{Modelling dark energy}\label{sec2}


\subsection{Quintessence models}

In this work, we consider the simplest class of quintessence models (\cite{quint1}; \cite{quint2})
in which a scalar field $Q$ is slow-rolling in a runaway potential. Numerous forms of potentials
have been proposed but we restrict to two classes of potentials. Let us briefly summarize the
properties of these models.

The first class of potentials is an inverse power law (\cite{quint1}; \cite{quint2})
\begin{equation}\label{eq:RP}
 V(Q)=M^4(Q/M_{\rm p})^{-\alpha},
\end{equation}
often quoted as Ratra-Peebles (RP) potential. $M_{\rm p}\equiv(8\pi G)^{-1/2}$ is the
reduced Planck mass. The potential depends on two free parameters: $\alpha$ is a positive
index while $M$ is a mass scale that has to be adjusted to fit $\Omega_\q$ today, once
$\alpha$ is given. In particular, if $\Omega_\q$ dominates today then $M$ and $\alpha$ are
related by
\begin{equation}
 \log\left(\frac{M}{1\,{\rm
 GeV}}\right)\sim\frac{19\alpha-47}{4(\alpha+1)}.
\label{mass}
\end{equation}
The second class of potential is an extension of the previous that takes supergravity
(SUGRA) corrections into account when $Q\sim M_{\rm p}$ (\cite{braxmartin99}),
\begin{equation}\label{eq:SUGRA}
 V(Q)=M^4(Q/M_{\rm p})^{-\alpha}\exp(Q^2/2M_{\rm p}^2).
\end{equation}
Both potentials have a similar dynamics as long as $Q\ll M_{\rm p}$ but differs at low
redshift, in particular concerning their equation of state. In the SUGRA case, it is pushed
toward $-1$ and one expects $w_0\sim-0.82$ (\cite{braxmartin99}).

\begin{figure}[b]
 \centering
   \includegraphics[width=8.8cm]{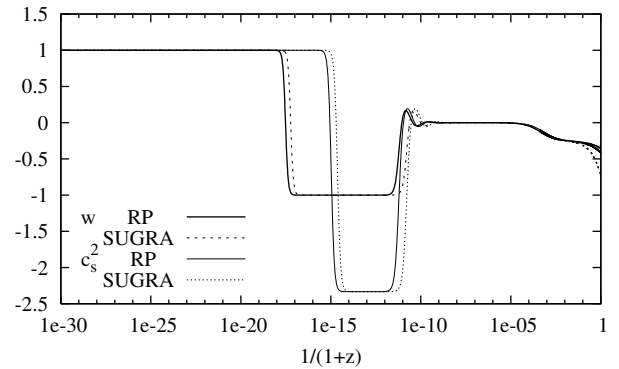}
 \caption{Evolution of the equation of state $w$ and of the sound speed $c_s$ with
 the redshift for an inverse power law quintessence model with $\alpha=6$, including or
 not the supergravity correction. We recover, from high to low redshift,
 the kinetic, slow-rolling and tracking phases described in the text.}
 \label{fig:2}
\end{figure}

With such well-defined models, the dynamics of the background is completely characterized by
the Klein-Gordon equation for the scalar field,
\begin{equation}
 \ddot Q + 3H\dot Q+ \frac{\dd V}{\dd Q}=0,
\end{equation}
in addition to the Friedmann equation
\begin{equation}
 \left(H^2+\frac{K}{a^2}\right) = \frac{8\pi G}{3}\left[\rho_\mat + \rho_\rad +
 \frac{\dot Q^{2}}{2} + V(Q)\right]
\end{equation}
allowing for this new matter contribution. These equations characterize background and low
redshift observations and in particular the linear growth factor. Let us stress that, at
this level of description, one can describe the quintessence component as a fluid with a
time-dependent equation of state. This is due to the fact that the speed of sound,
$c_s^2$, is given by
\begin{equation}\label{eq:csQ}
 c_s^2=1+\frac{4}{3}\frac{1}{H\dot Q}\frac{\dd V}{\dd Q}
\end{equation}
and that the equation of state evolves as
\begin{equation}\label{eq:wpcs}
 \dot w=-3H(1+w)(c_s^2-w).
\end{equation}
These models share the interesting property to possess scaling solutions which are attractor
of the dynamical evolution. In general, but depending on the initial conditions, the
dynamics starts with an early kinetic phase ($\dot Q^2\gg V$) in which $w\sim1$ so that
$\rho_\q\propto(1+z)^6$. The field behaves as $Q=Q_i-A(1+z)$ and it freezes to a constant
value. Since the kinetic energy decreases while the potential remains constant, this regime
cannot last forever. When the potential starts dominating, the equation of state shifts
suddenly to $w\sim-1$. During this transition regime, $\rho_\q\propto(1+z)^0$. Then, there
is a potential regime that lasts until the tracking regime during which
\begin{equation}
 w=c_s^2=\frac{\alpha w_\mathrm{B} -2}{\alpha+2}
\end{equation}
where $w_\mathrm{B}$ stands for the equation of state of the fluid dominating the background. At
that stage, the scalar field is slow-rolling so that $w<0$. Fig.~\ref{fig:2} depicts the evolution
of $w$ and $c_s^2$ during these various regimes.

When cosmic microwave background anisotropies and large scale structures are considered, one
needs to include the description of the evolution of the perturbations, and in particular
include those of the scalar field described by
\begin{equation}\label{eq:qpert}
 \delta\ddot Q + 3H\delta\dot Q + \left(\frac{k^2}{a^2}+\frac{\dd^2 V}{\dd
 Q^2}\right)\delta Q + \mathcal{S}=0
\end{equation}
where $\mathcal{S}$ encodes the perturbations of the metric of the spacetime and $k$ is the
comoving wavenumber of the perturbation. It was shown (\cite{braxetal}; \cite{cmbslow}) that there
exists an attraction mechanism for super-Hubble wavelength so that the spectrum is
insensitive to the initial conditions for the scalar field.


\subsection{Models and parameterizations}\label{subsec22}

Most data, and in particular supernovae data, are being analyzed using a general
parameterization of the equation of state. These parameterizations are useful to extract
model-independent information from the observations but the interpretation of these
parameters is not always straightforward. In this paragraph, we remind the properties of
some interesting parameterizations and compare them to the quintessence models we are
considering.

Let us first recall that general parameterizations of the equation of state as
\begin{equation}\label{eq:para4}
w(a)=w(a_0) + [w(a_m)-w(a_0)]\ \Gamma(a,a_t,\Delta)
\end{equation}
were shown to allow an adequate treatment of a large class of quintessence models
(\cite{para1}; \cite{para2}). Such a parameterization involves four parameters $\lbrace
w(a_0),w(a_m),a_t,\Delta\rbrace$ and a free function $\Gamma$ varying smoothly between one
at high redshift to zero today. Even though it reproduces the equation of state of most
quintessence models, it is not economical in terms of number of parameters since most
quintessence potentials involve one or two free parameters. If one assumes that the
parameterization is supposed to describe the dynamics of a minimally coupled scalar field,
the knowledge of $w$ is sufficient but in a more general case one would need more information:
The background dynamics depends on the potential and its first derivative, which can be related
to $w$ and $\dot w$. Accounting for perturbations, one needs to know the second derivative
of the potential [see Eq.~(\ref{eq:qpert})] which can be inferred from $\ddot w$ (\cite{dave}).

\begin{figure}[bht]
 \centering
   \includegraphics[width=8.8cm]{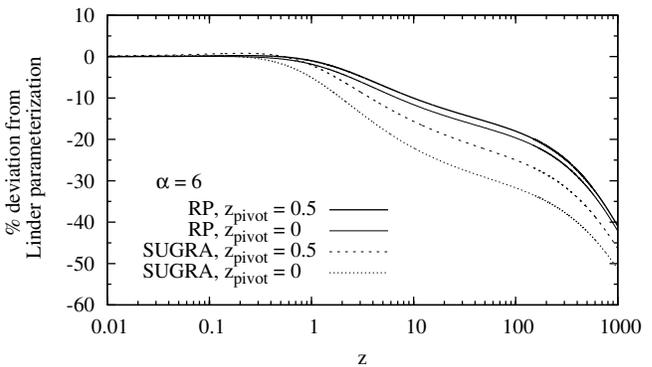}
 \caption{Deviation of quintessence equation of state for Ratra-Peebles (solid) and SUGRA
  (dotted) models with $\alpha=6$ from the generalized parameterization, Eq.~(\ref{eq:para4}),
  setting $z_\pivot = 0$ (thick) or $z_\pivot = 0.5$ (thin). Fitting the previous one up to
  $z\lesssim 0.3$, a deviation larger than 2\% occurs at $z\simeq 1$ for Ratra-Peebles models
  while at $z\simeq 0.5$ for SUGRA models.}
\label{fig:2a}
\end{figure}

\begin{figure*}[bht]
 \centering
   \includegraphics[width=12cm]{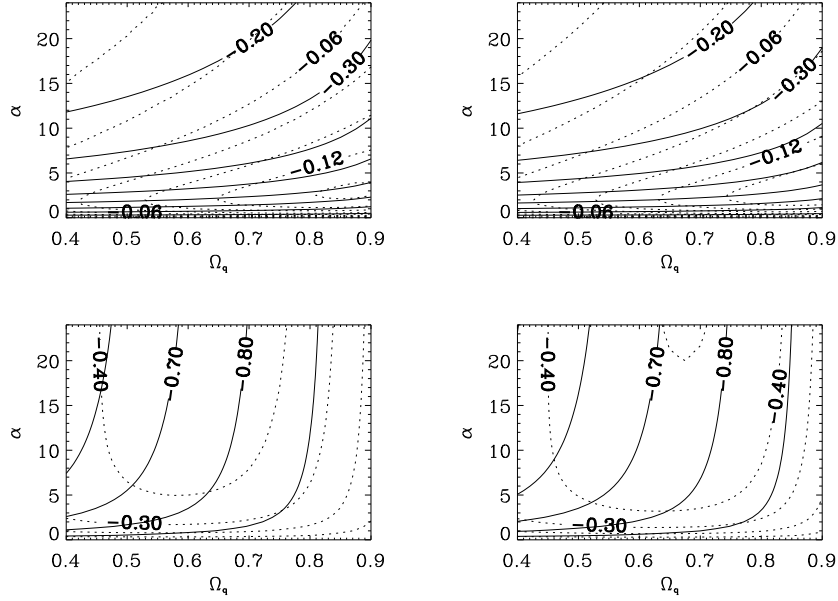}
 \caption{Contour plots of the quintessence equation of state. We compare the equation
 of state of two quintessence models with the parameterization~(\ref{eq:genLinder}) for
 two values of the pivot redshift:  $z_\pivot=0$ (left) and $z_\pivot=0.5$ (right). Solid
 lines correspond to level contours for $w_\pivot$ while dotted lines correspond to level
 contours of $w_a$. We have chosen the spacing of all the contour lines to be $\Delta w=0.1$,
 except for the plots in the upper line, where $\Delta w_a=0.02$. The upper line corresponds
 to Ratra-Peebles models, Eq.~(\ref{eq:RP}), while the lower line corresponds to SUGRA
 models, Eq.~(\ref{eq:SUGRA}). Due to the exponential correction, $w_0$ is always smaller
 for SUGRA models because the potential is flatter and the field is rolling slower. Also,
 the value of $w_\pivot$ and $w_a$ are more sensitive to the choice of $z_\pivot$ for SUGRA
 models than for Ratra-Peebles models.} \label{fig:2b}
\end{figure*}

Since we expect dark energy to have observable consequences on the dynamics only at late
time, one can consider an equation of state obtained as a Taylor expansion around a pivot
point,
\begin{equation}\label{eq:genLinder}
 w(a) = w_\pivot + w_a(a_\pivot - a).
\end{equation}
This form depends on only two parameters and is a generalization of the parameterization
proposed by~\cite{para3} and then \cite{para3bis} where $a_\pivot=1$. Two considerations are
in order when using such a parameterization. First, the redshift band on which this is a
good approximation of the equation of state is unknown. Clearly, compared with the
form~(\ref{eq:para4}), it is unlikely to describe dark energy up to recombination time; see
Fig.~\ref{fig:2a}. Secondly, when combining observables at different redshift such as weak
lensing, Sn~Ia and CMB, one should choose the value of $a_\pivot$ in such a way that the
errors in $w_\pivot$ and $w_a$ are uncorrelated (\cite{hujain}). It follows that the pivot
redshift is the redshift at which $w$ is best constrained. In particular it was argued that
it is important to choose $a_\pivot\not=1$ for distance-based measurements. The problem lies
in the fact that the pivot redshift is specific to the observable. In this respect, dark
energy models defined by a Lagrangian are more suitable, yielding to a definite equation of
state as a function of redshift, hence more general than a Taylor expansion around a pivot
point. Eventually, one can read out the values of $w_\pivot$ and $w_\mathrm{a}$ at whatever
redshift. Fig.~\ref{fig:2b} depicts the value of $w_\pivot$ and $w_a$ for the quintessence
models we use, Eqs.~(\ref{eq:RP}) and~(\ref{eq:SUGRA}), for the pivot redshifts $z_\pivot=0$
and $z_\pivot=0.5$. This complication, arising when one wants to combine datasets with
different $z_\pivot$, will also make it more difficult to infer constraints on the physical
models from the constraints on the parameterization.

There is an alternative way to get a first hint on the nature of dark energy. It may be
useful to consider the plane $(w,w')$ where $w'\equiv\dd w/\dd \ln a$ is the derivative of
$w$ with respect to the number of $e$-folds. It was recently shown by Caldwell \& Linder (2005)
and Scherrer (2005) that quintessence models occupy a narrow part of this plane. This can be
understood from Eq.~(\ref{eq:wpcs}) which implies that $w'+3(1-w^2)=3(1+w)(1-c_s^2)$. For
quintessence models, $1+w>0$, and Eq.~(\ref{eq:csQ}) implies that $c_s^2<1$ (because
$\dot Q>0$ and $V'<0$) so that
\begin{equation}\label{eq:wpw1}
 w'>-3(1-w^2),
\end{equation}
without any assumptions on the dynamics of the scalar field. In Caldwell \& Linder (2005),
two classes of quintessence models where exhibited, namely ``thawing'' models, in which $w\sim-1$
initially and increases as $Q$ rolls down the potential , and ``freezing'' models, in which
$w>-1$ initially and tends toward $-1$ as $Q$ rolls down the potential. ``Freezing'' models
contain tracking models and in particular the Ratra-Peebles models, Eq.~(\ref{eq:RP}), and
SUGRA models, Eq.~(\ref{eq:SUGRA}), considered in this work. Using a combination of
numerical simulations and physical arguments, they concluded that
\begin{equation}\label{eq:wpw2}
 3w(1+w)<w'<0.2w(1+w)
\end{equation}
for ``freezing'' models. From an observational point of view, the analysis of supernovae
data~(\cite{rsn1w}) showed that if the universe is flat then $w_0=-1.31^{+0.22}_{-0.28}$ and
$w'_0=-1.48^{+0.90}_{-0.81}$ after marginalizing on $\Omega_{\mat0}$. If one further imposes
that $w_0>-1$ then $w_0<-0.76$ and $w'_0=-0.6\pm0.5$. In the case where $w$ is assumed
constant then $w=-1.02^{+0.13}_{-0.19}$ and $w<-0.72$ at 68\% and 99\% confidence level
respectively.

\begin{figure}[hbt]
 \centering
   \includegraphics[width=7cm]{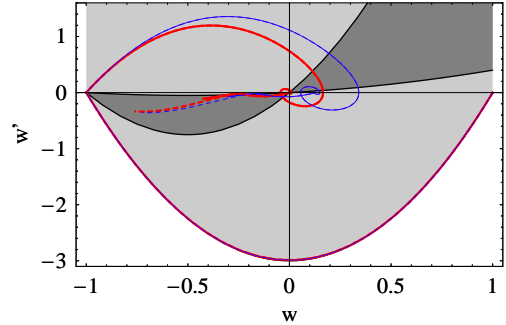}
  \caption{Dynamics of the two quintessence models in the  plane $(w,w')$.
  The shaded regions correspond to the constraints~(\ref{eq:wpw1})  in light
  gray and~(\ref{eq:wpw2}) in dark grey. We have considered a Ratra-Peebles
  (solid) and SUGRA (dash) models with $\alpha=6$ (thick/red) and $\alpha=11$
  (thin/blue). Only in the tracking regime the models are compatible with
  Eq.~(\ref{eq:wpw2}).}
\label{fig:wwp}
\end{figure}

This phase-space analysis creates a link between the various parameterizations and the physical
models. It can be shown~(\cite{wwprim2}) that different classes of models (e.g. k-essence,
Chaplygin gaz, quintessence, etc.) lies in different parts, hence offering a way to distinguish
between these models without measuring $w(z)$. In particular, Fig.~\ref{fig:wwp} depicts the
dynamics of some Ratra-Peebles and SUGRA models in the $(w,w')$ plane, superposed to the regions
where the inequalities~(\ref{eq:wpw1}) and~(\ref{eq:wpw2}) hold. Notice that the trajectories for
the SUGRA potential are essentially the same of the Ratra-Peebles one, deviating just at low
redshift towards the cosmological constant solution.


\subsection{Summary}

Quintessence models require only {\it two} parameters to describe the whole dynamics (with no
redshift limitation). Compared to a pure cosmological constant, described by only one number, this
gives us one extra parameter.  In terms of extra-parameter with respect to a standard
$\Lambda$CDM, this is equivalent as considering a constant equation of state for dark energy.

On the other hand, a parameterization of the equation of state of the dark energy is
sufficient to describe the low redshift universe. But, the parameterizations which describe
accurately quintessence models involve at least \emph{four} extra-parameters. It is thus
more economical to work directly with the physical model. In that case, the evolution of
perturbations can be inferred from $w$ but this in not the case in more general situations.

Parameterizations with fewer parameters have a smaller dynamical range and are likely to be
bad descriptions of dark energy at high redshift, notably for CMB computation. Besides, when
combining dataset, the choice of the pivot redshift may induce some difficulties in
interpreting the constraints on the parameters in terms of constraints on the physical
models.

So considering directly a well-defined physical model instead of a parameterization is
economical concerning the number of extra-parameter and avoids the problem of the pivot
redshift. It allows us to compute the prediction of the models at all redshift (Sn~Ia, weak
lensing, and CMB). On the other hand, it concerns only a small class of models.

This discussion shows that both routes are complementary. In particular, it would be worth
to evaluate to which accuracy constraints of order 1\% on a given parameterization constrain
physical models.


\section{Cosmic shear}\label{sec3}

Gravitational lensing by large scale structures of the universe produce  weak distortion 
fields and collectively modify the shape of background galaxies (see e.g.~\cite{bartelmann01};
\cite{mellier99}; \cite{refregier03}). Though this is a very weak signal and a challenging task,
it has been  detected  almost simultaneously by van Waerbeke \etal (2001), Wittman \etal (2000),
Kaiser \etal (2000), Bacon \etal (2000) and it now routely  observed by many groups around the
world (see \cite{VWM03}; \cite{HOEK03b} for recent reviews on observations). Over the past years
the huge  efforts carried out by these groups to deal with critical systematics considerably
improved the reliablity of the lensing signal  and stenghtened the ability of cosmic shear to
constrain cosmological models from the statistical analyses of galaxy ellipticies.


\subsection{Generalities}\label{sec3a}

The gravitational lensing effect depends on the second order derivatives of the
gravitational potential. The convergence, $\kappa$, and the shear,
$\bm{\gamma}=(\gamma_1,\gamma_2)$, describe the distortion of background images by the
matter along the line of sight. These components are related by
\begin{equation}
 \Delta\kappa=(\partial_1^2-\partial_2^2)\gamma_1 +
 2\partial_1\partial_2\gamma_2.
\end{equation}
The evolution of the convergence is dictated by the Sachs equation (\cite{sachs62}; see~\cite{ub00}
for a more modern description). The convergence in the direction $\bm{\theta}$ can be related to
the matter distribution integrated along the line of sight
\begin{eqnarray}\label{eq:kap}
 \kappa(\bm{\theta},\chi) &=& \frac{3}{2}\frac{\Omega_{\mat0}H_0^2}{c^2}
 \int_0^{\chi}\frac{S_K(\chi-\chi')S_K(\chi')}{S_K(\chi)}\nonumber\\
 && \qquad\qquad\qquad\quad \times\frac{\delta_\mat[S_K(\chi')\bm{\theta},\chi']}{a(\chi')}\dd\chi'
\end{eqnarray}
for sources located at a radial distance $\chi$ defined by
\begin{equation}
 \chi(z)=\int_0^z\frac{\dd z'}{H(z')}
\end{equation}
and $S_K$ is the angular diameter distance given by
\begin{equation}
 S_K(\chi) = \left\lbrace
 \begin{array}{lc}
   \sin(\sqrt{K}\chi)/\sqrt{K} & K>0 \\
    \chi & K=0 \\
   \sinh(\sqrt{-K}\chi)/\sqrt{-K} & K<0\\
 \end{array}
 \right..
\end{equation}
If the sources have a distribution given by
$n_\chi(\chi)\dd\chi=n(z)\dd z$ then the effective convergence
takes the form
\begin{equation}
  \kappa(\bm{\theta})=\int_0^{\chi_H}
  n_\chi(\chi)\kappa(\bm{\theta},\chi)\dd\chi
\end{equation}
where $\chi_H$ is the comoving radial distance of the horizon. Decomposing the convergence in
2-dimensional Fourier modes,
\begin{equation}
 \kappa(\bm{\theta}) = \int
 \frac{\dd^2\bm{\ell}}{2\pi}\hat\kappa(\bm{\ell})\,\hbox{e}^{i\bm{\ell}\cdot\bm{\theta}},
\end{equation}
the shear power spectrum, defined by $\langle\hat\kappa(\bm{\ell}) \hat\kappa(\bm{\ell}')\rangle =
P_\kappa(\ell)\delta^{(2)} (\bm{\ell}+\bm{\ell}')$, can be related to the 3-dimensional power
spectrum of matter density perturbations $P_\mat$ by
\begin{eqnarray}\label{eq:pkappa}
 P_\kappa(\ell) = \frac{9}{4} \frac{H_0^4\Omega_{\mat0}^2}{c^4}\int_0^{\chi_H}
 \left[\frac{g(\chi)}{a(\chi)}\right]^2P_\mat\left[\frac{S_K(\chi)}{\ell},\chi \right]\dd\chi
\end{eqnarray}
in the small angle approximation (see
e.g. \cite{bartelmann01}, \cite{pubook} chap.~7). The
function $g$ is given by
\begin{equation}
 g(\chi) = \int_\chi^{\chi_H}
 n_\chi(\chi')\frac{S_K(\chi'-\chi)}{S_K(\chi')}\dd\chi'.
\end{equation}
Note that the window function $W(z)\equiv [g(\chi)/a(\chi)]^2$ is  peaked around $z\simeq z^*/2$
for a distribution of sources $n(z)$ approximately peaked at redshift $z^*$. This will be useful
in the choice of the pivot redshift. Let us stress that expressions~(\ref{eq:kap})
and~(\ref{eq:pkappa}) assume the validity of the Poisson equation and thus of general relativity.
These expressions may be slightly different in more general contexts and even be used to test
general relativity (see~\cite{ubpoisson}; \cite{SUR04}).

Neither $\kappa$ nor $P_\kappa$ are directly observable, but only filtered quantities can be
obtained. Cosmic shear can be measured by various types of 2-point statistics which differ only by
the chosen filtering scheme. This implies that their sensitivity to the power spectrum, and also
to systematics, are different. In this work we consider two of these statistics, namely the
aperture map variance, defined by
\begin{equation}\label{eq:map2}
 \langle M_\ap^2\rangle(\theta_c) = \frac{288}{\pi}\int \ell
 P_\kappa(\ell)\left[\frac{J_4(\ell\theta_c)}{\ell^2\theta_c^2}\right]^2\dd\ell \, ,
\end{equation}
which is a bandpass estimate of the convergence power spectrum, and the top-hat shear variance
\begin{equation}\label{eq:thsv}
 \langle \gamma^2\rangle(\theta_c) = \frac{8}{\pi}\int \ell
 P_\kappa(\ell)\left[\frac{J_1(\ell\theta_c)}{\ell\theta_c}\right]^2\dd\ell \, ,
\end{equation}
which is a lowpass estimate of $P_\kappa$. Here $J_n$ are the Bessel functions of the first kind.
Both statistics can be deduced from two linear combinations of the radial and tangential
components of the shear variance, $\xi_\pm \equiv \langle\gamma_{\rm{t}}^2\rangle \pm
\langle\gamma_{\rm{r}}^2\rangle$ (see e.g. \cite{bartelmann01}), which are directly estimated from
the shapes of background galaxies.


\subsection{Matter power spectrum and non-linear regime}

In the previous expressions, and in particular in Eqs.~(\ref{eq:kap}) and~(\ref{eq:pkappa}),
$P_\mat$ refers to the full 3-dimensional power spectrum of pressureless matter, including cold
dark matter and baryons.

In the linear regime, the growth factor $D_+$ is the growing solution of
\begin{equation}
 \ddot D_+ + 2 H\dot D_+ -\frac{3}{2}H^2\Omega_{\mat}D_+ = 0.
\end{equation}
The second term that describes the damping due to the cosmological expansion contains all
the effect of dark energy on $D_+$. In this regime, the total effect on lensing,
Eq.~(\ref{eq:pkappa}), is similar to the one obtained from a single redshift plane
(see~\cite{ludoben}; \cite{SUR04}) so that the integrated growth effect is degenerated with the
normalization of the spectrum.

Dealing with low-redshift sources, the lensing predictions and analysis involve the
non-linear power spectrum. This regime cannot be described analytically from a perturbation
approach (see however the proposal by~\cite{Scoccimarro05}) and one would need to rely on
$N$-body simulations.

$N$-body simulations including quintessence were recently performed using the GADGET code
(\cite{Nbodyquint}) or an adaptive refinement tree code (\cite{Nbodyquint2}). It was argued
by Dolag \etal (2003) that the halo concentration distribution around the mean value does not
depend on the cosmology, while the concentration parameter depends on the dark energy equation of
state at the cluster formation redshift through the linear growth factor. However a systematic
study confirming this claim is lacking. Klipin \etal (2003) and \cite{Nbodyquint3} show that dark
energy changes the virial density contrast, $\Delta_c$, which induces a change in the power
spectrum at small scales ($k\ga 1h\,\mathrm{Mpc}^{-1}$) but, for constant $w$, the error is
smaller than the error in the expected non-linear model (\cite{jjbd}).

Instead of specific $N$-body simulations accomodating quintessence, in order to compute the
non-linear matter power spectrum one can deal with linear--to--non-linear mappings, for
instance based on the stable clustering {\it ansatz} (\cite{hamilton}; \cite{peacockd}) or
to a halo model (e.g. \cite{lnlmafry}; \cite{lnlseljak}; \cite{smithetal}). These mappings have
been tested for several cosmologies including $\Lambda$CDM but not for dynamical dark energy models.
Given their robustness and the precision level we can actually reach, we can hope that they remain
valid for the class of models we are considering. Indeed this is a very strong assumption that can
be justified by the fact that we do not expect the scalar field to cluster on small scales
so that it is unlikely to affect the small scale behaviour of matter but by its influence on
the expansion rate. Arguing that the clustering scale of the quintessence field is given by
its Compton wavelength, Ma \etal (1999) propose an analytic approximation for
the Peacock \& Dodds (1996) formula to include quintessence with constant equation of state,
claiming a 10\% level accuracy. \cite{lnlquint2} propose a recipe to extend the
aforementioned mappings to $w\neq -1$ cases, attaining a better accuracy for $k\la
10$~Mpc$^{-1}$ by systematically exploring a wider parameter space. Nevertheless, upcoming
weak lensing measurements require an improved description of the smaller scales physics
(\cite{NLrequired}), eventually including hydrodynamics (\cite{baryons}; \cite{baryons2}).

In conclusion, it seems early to decide how dark energy, and quintessence in particular,
modify the mapping calibrated on $\Lambda$CDM. For that reason, in this work we will consider
two linear--to--non-linear mappings, by Peacock \& Dodds (1996) and Smith \etal (2003), and try
to identify the parameters that are not sensitive to this choice. We will also try to quantify
how the other parameters are affected so that we can estimate how our ignorance of the non-linear
regime limits the use of weak lensing.

Hopefully, as we will also show, weak lensing data in the linear regime shall be able to be
used. In that case, we can get an interesting constraint on the growth factor without
messing with non-linear physics.

\begin{figure*}[htb]
 \centering
   \includegraphics[angle=90,width=14cm]{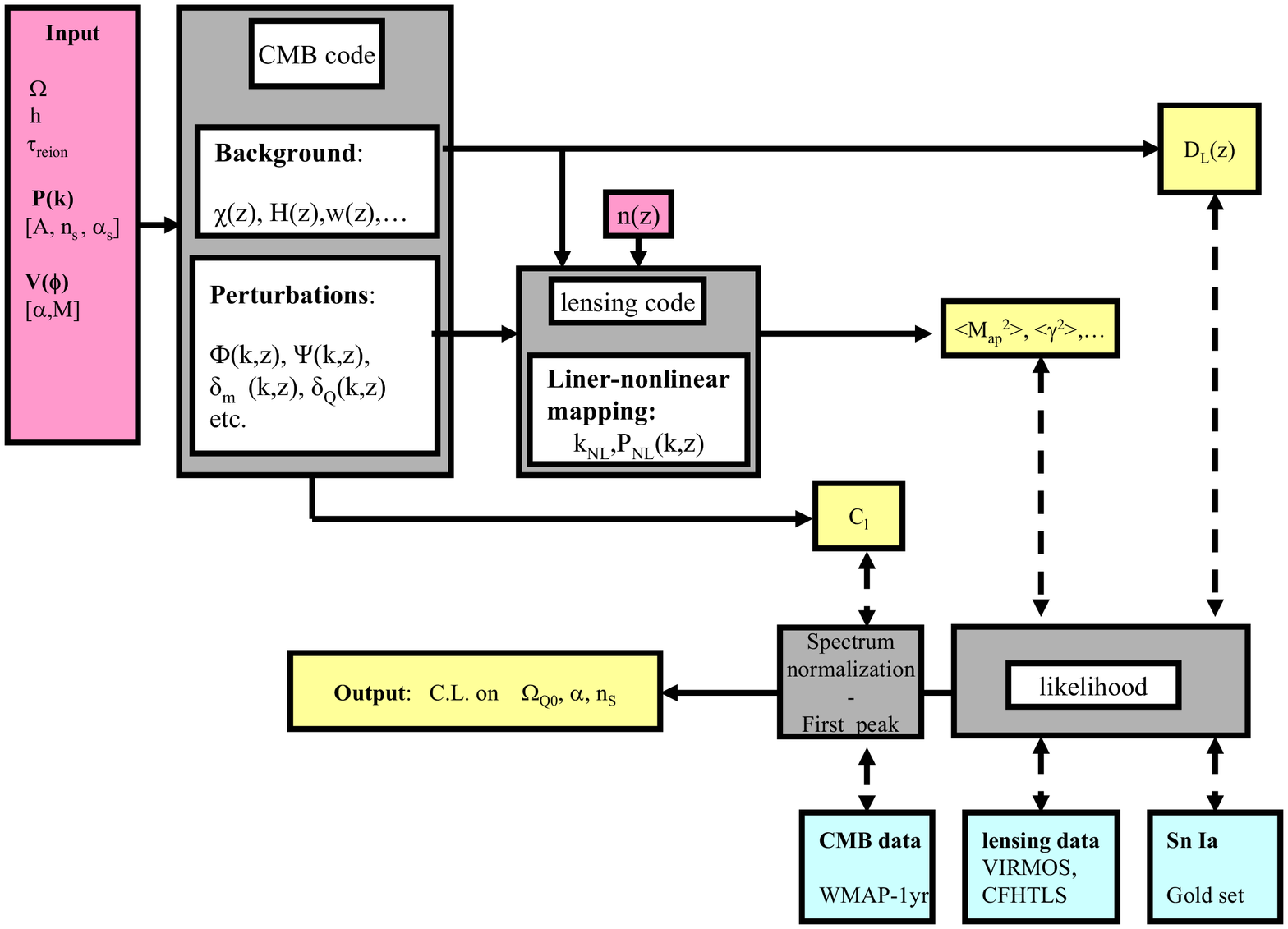}
 \caption{Pipeline implemented for this work. Presently, we restrict to three free cosmological
 parameters, $\{\Omega_{\q 0},\alpha, n_s \}$, keeping fixed the others. The lensing code
 manipulates both background's and perturbations quantities computed by the CMB code, using a
 sources distribution $n(z)$ depending on the used dataset. In particular, the source redshift
 parameter $z_\mathrm{s}$ is left to vary and marginalized over afterwards. Finally, the
 likelihood is computed using (either real or synthetic) cosmic shear and Sn~Ia data, both
 separately and jointly. The temperature CMB data are used to fix the amplitude $A$ of the power
 spectrum at decoupling, and to put (conservative) constraints on the $(\Omega_{\q 0},\alpha)$
 parameters sub-space using the location of the first peak. In the CMB section of the pipeline,
 we indicate by $\delta_\mat$ and $\delta_\q$ energy density fluctuations in matter and
 quintessence components, respectively, while $\Phi(k,z)$ and $\Psi(k,z)$ are the scalar
 perturbations of the metric (Bardeen potentials) in Fourier space. In the lensing section, we
 denote by $k_\mathrm{NL}$ the scale at which the power spectrum becomes non-linear,
 $P(k_\mathrm{NL},z)\sim 1$. See \S~\ref{sec3} for details.}
 \label{fig:chart1}
\end{figure*}


\subsection{Lensing data}\label{sec3b}

We use three sets of data for weak lensing, the VIRMOS-Descart (\cite{VIRMOSDescart05}),
the deep field of the CFHTLS survey (\cite{CFHTLSdeep05}) and the wide field of CFHTLS
(\cite{CFHTLSwide05}).

The details regarding the survey properties, image and catalogue processing  of VIRMOS-Descart
data are described in \cite{vw01} and \cite{mccrackenetal03}. The shear measurement and error
analysis are described in \cite{VIRMOSDescart05} and summarized in Table 2 of that paper. This
survey covers an effective (the unmasked area) area of $8.5$~deg$^2$ spread in four fields. It
probes lensed galaxies down to the limiting magnitude $I_{AB}=24.5$ has an effective galaxy number
density (after all selection processes) of 15 gal/arcmin$^2$, and explores angular scales up to 50
arc-minutes.

For the CFHTLS deep and wide, all relevant details are given in \cite{CFHTLSdeep05} and
\cite{CFHTLSwide05}, respectively. The deep covers an effective area of $2.2$~deg$^2$ in three
fields, down to $I_{AB}=26$, but only samples angular scales up to 30 arc-minutes. The effective
galaxy number density is $22$~gal/arcmin$^2$. In contrast, the wide has a much larger effective
area than the deep ($22$~deg$^2$) but only spread in two fields. It explores angular scales up to
one degree at about the same depth as VIRMOS-Descart. The effective galaxy number density is
$13$~gal/arcmin$^2$ (the final selection produced a catalogue somewhat less deep than
VIRMOS-Descart). Note that once completed the wide survey will be composed of three compact areas
that will sample angular scales up to 5 degrees in three independent fields so that linear scales
will be explored with much more accuracy than present-day CFHTLS wide data.

Each survey used complementary photometric or spectroscopic galaxy samples to derive the redshift
distribution of the lensed galaxy samples. As discussed in the VIRMOS-Descart and CFHTLS cosmic
shear papers, it is convenient to describe the redshift distribution by a three-parameter
 function
\begin{equation}\label{eq:sourcedist}
 n(z)=\frac{\beta}{z_\mathrm{s}\Gamma\left(\frac{1+\alpha}{\beta}\right)}
 \left(\frac{z}{z_\mathrm{s}}\right)^\alpha\exp\left[-\left(\frac{z}{z_\mathrm{s}}\right)^\beta
 \right].
\end{equation}
$\alpha$ and $\beta$  are obtained from a fit of photometric redshift distributions derived from
external redshift calibration surveys. The CFHTLS and VIRMOS are different lensing data
sets, but they were obtained with the same telescope (although with different instruments) and with
the same exposure time. Therefore the two lensing surveys have to be calibrated using the same external
redshift data set in order to preserve the homogeneity of the analysis. In van Waerbeke \etal (2005),
VIRMOS was calibrated using the Hubble Deep Fields and MS1008 (see van Waerbeke \etal 2001 for the
details), while the CFHTLS lensing data (Hoekstra \etal 2006; Semboloni \etal 2006) were calibrated
using the HDF only. The HDF and MS1008 data in visible and near infrared bands produced accurate
photometric redshifts (Yahata \etal 2000; Athreya \etal 2002).
We decided to use the HDF only for both VIRMOS and CFHTLS and abandon the MS1008 field. We have checked
that the results discussed in this paper do not depend whether or not we include the MS1008 calibration
field, since the sample variance due to the use of combined redshift calibration sets is absorbed in
the redshift error (van Waerbeke \etal 2006).

In the following, we will use for the CFHTLS-deep survey (\cite{CFHTLSdeep05})
\begin{equation}\label{eq:nzdeep05}
 \alpha=1.9833,\quad \beta=0.6651,\quad
 z_\mathrm{s}=0.0981^{+0.0129}_{-0.0114}{}^{+0.0209}_{-0.0161}
\end{equation}
giving a mean redshift $\langle z \rangle= 1.01$, while for the CFHTLS-wide and the VIRMOS-Descart
surveys (\cite{CFHTLSwide05})
\begin{equation}\label{eq:nzwide05}
 \alpha=1.35,\quad \beta=1.654,\quad
 z_\mathrm{s}=0.668^{+0.035}_{-0.036}{}^{+0.053}_{-0.055}
\end{equation}
giving a mean redshift $\langle z \rangle= 0.76$. We quote the $1\sigma$ and $2\sigma$ errors. We
use the same parameters for the  VIRMOS-Descart and CFHTLS-wide redshift distributions since both
surveys have a similar depth and the same effective galaxy number density.

\begin{figure*}[thb]
 \centering
   \includegraphics[width=17cm]{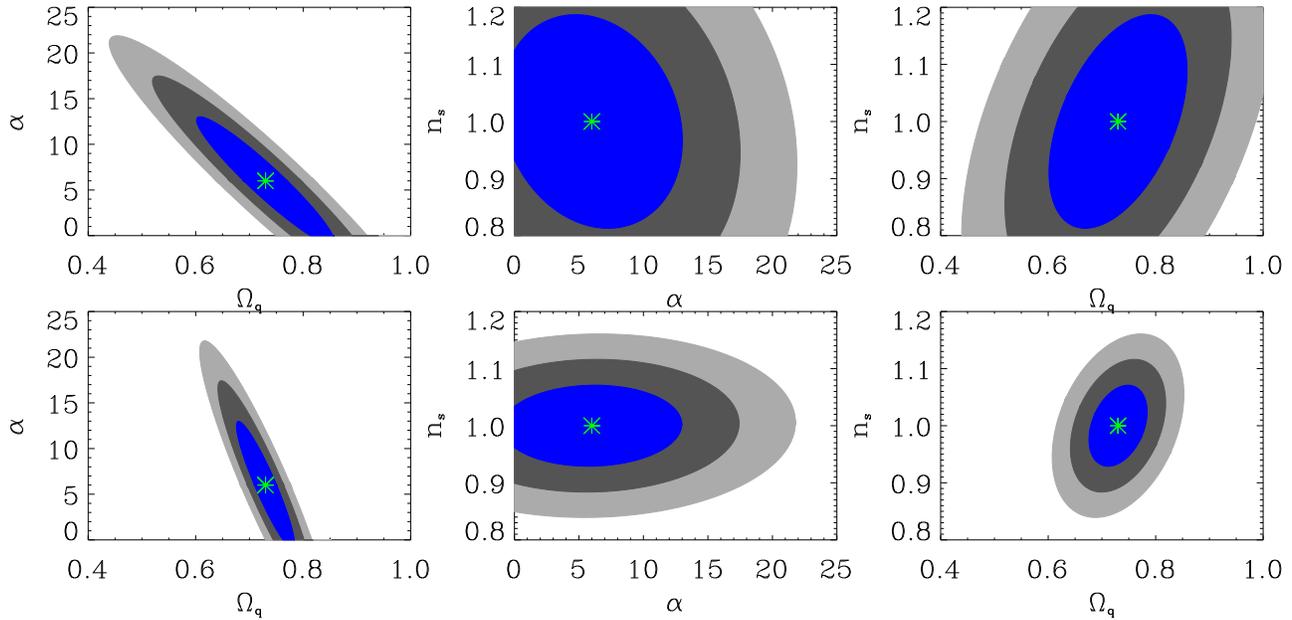}
 \caption{Fisher analysis of the top-hat shear variance of the cosmic shear on
 cosmological parameters $(\alpha,\Omega_{\q 0},n_s)$, for Ratra-Peebles (top line)
 and SUGRA (bottom line) models  -- contours at 68 and 95\% C.L. We employ a
 synthetic realization of a CFHTLS-wide like survey, covering 170~deg$^2$, with
 $20$~gal/arcmin$^2$ and intrinsic ellipticity $\sigma_e=0.4$. The fiducial model is
 defined by $(\alpha,\Omega_{\q 0},n_s)=(6,0.73,1)$ while $(h,\tau_\mathrm{reion},
 \Omega_{\mathrm{b} 0}h^2)=(0.72,0.17,0.024)$ are kept fixed. The goal of this analysis
 is to help in choosing the range of the grid for the likelihood analysis. See \S~\ref{sec3c}
 for discussion.}
 \label{fig:3afisher}
\end{figure*}


\subsection{Description of the pipeline and choice of the free parameters}\label{sec3c}

The pipeline we developed is summarized in Fig.~\ref{fig:chart1}.
We compute the evolution of background and perturbations power spectra in linear regime by
means of a Boltzmann code allowing for photons, neutrinos, baryons, cold dark matter and
quintessence scalar field. Notice that this code deals with several gauge choices and can
also account for scalar-tensor theories of gravity (\cite{cmbslow}), hence allowing to study
deviations from general relativity in this framework as well as extended quintessence
scenarios (\cite{jpu99}; \cite{amendola}; \cite{chiba}).

Using this code we compute the CMB temperature (TT) angular power spectrum ($C_\ell$) in order to
fix the amplitude of the initial matter power spectrum at the redshift of the last scattering. We
do it by matching the computed $C_\ell$ with WMAP-1yr data (Hinshaw \etal 2003) at a high
multipole, to be preferred when studying quintessence since at low multipoles the integrated
Sachs-Wolfe effect is dominant so that temperature anisotropies are not directly related to the
density perturbations. Definitely, we use the multipole $\ell\simeq 110$ of WMAP-1yr data, where
the total relative error of the TT spectrum is smaller than 3\%. This normalization procedure
holds until the correlation between multipoles is weak and does not take into account the
measurement errors on the amplitude of the TT spectrum. However, we expect that the final results
would not be strongly affected by a more accurate normalization. Notice that, as mentioned
earlier on, $\sigma_8$ data is not used to normalize the spectrum. Hence, its value may be
evaluatd from the matter power spectrum and compared with the observed values.

Once the linear matter power spectrum is known at every redshift, the weak lensing add-on
code (\cite{SUR04}) computes the non-linear power spectrum using two mappings, by
Peacock \& Dodds (1996) and Smith \etal (2003). We consider wavevectors ranging up to $10\,h\mbox{
Mpc}^{-1}$ (corresponding to 1 arcmin at $z \sim 1$ for cosmologies close to $\Lambda$CDM).
By Limber projection, we deduce the shear power spectrum allowing for a source redshift
distribution of the form~(\ref{eq:sourcedist}). To finish, several two-point statistics in
real space are computed, namely the top-hat shear variance $\langle \gamma^2\rangle$, the
aperture mass variance $\langle M_\ap^2\rangle$, as well as the two-point correlations
$\xi_\pm$. Let us stress that, like the Boltzmann code, the lensing add-on code works with
scalar-tensor theories of gravity as well.

Cosmological parameters are estimated by comparing the predicted signal $m_i$ to the data $d_i$ as
a function of scale $X_i$ (which reduces to an angular scale $\theta_i$, a redhift $z_i$ or a
multipole $\ell_i$ respectively for lensing, supernovae and CMB data). We vary the parameters of
the model, disposing them on a regularly spaced grid and evaluate, at each grid point, the
likelihood function,
\begin{equation}
{\cal L}={1\over (2\pi)^n|C|^{1/2}} \exp\left[-\frac{1}{2}(d_i-m_i)^T{C}^{-1}(d_i-m_i)\right].
\label{like}
\end{equation}
Here $C^{-1}$ is the data covariance matrix, including Poisson shot noise and cosmic variance.

We focus on constraining the dark energy density and the parameter $\alpha$ of the quintessence
potentials and restrict to a low-dimensional parameter space. Ideally, one would include at least
nine cosmological parameters: the spatial curvature, the Hubble constant, the parameter $\alpha$
of the potential, the dark energy and matter (both dark matter and baryonic) density parameters,
the reionization optical depth, the amplitude and spectral index of the initial power spectrum.
In addition, one should include the three parameters accounting for the redshift distribution of
sources, Eq.~(\ref{eq:sourcedist}).

Given the result of the analysis of CMB data (\cite{wmap2}) we have assumed a spatially flat
universe so that $\Omega_{\mat0}=1-\Omega_{\q0}$. The amplitude of the initial power spectrum is
fixed by the normalization on the CMB. We have also assumed that the reduced Hubble constant, the
reionization optical depth, and the baryon energy density today are fixed to $h=0.72$,
$\tau_\mathrm{reion}=0.17$, and $\Omega_{\mathrm{b} 0}h^2= 0.024$, respectively.

We preliminary performed a Fisher matrix analysis on the parameters space $(\alpha,\Omega_{\q 0},
n_s)$ in order to estimate approximately the extent of the $1\sigma$ region and decide
the sampling steps of the grid for the computation of the likelihood. Figure~\ref{fig:3afisher}
depicts the 68\%, 95\%, and 99\% confidence levels (C.L.) after for cosmic shear (top-hat 
variance) corresponding to a synthetic CFHTLS-wide like survey covering 170~deg$^2$, with
20~galaxies/arcmin$^2$ and an intrinsic ellipticity of 0.4. The top line refers to Ratra-Peebles
models and the bottom line to SUGRA models. Interestingly, SUGRA models appear to be more
constrained. Notice that the Fisher analysis holds only locally, around a fiducial model (marked
by a cross) which is fixed at $\alpha=6$, $\Omega_{\q 0}=0.73$, and $n_s=1.0$.

In the final likelihood analysis we allow to vary the cosmological parameters in the
following ranges
\begin{equation}
 \qquad \Omega_{\rm{Q}0}\in [0.4,0.9],\quad
 \alpha\in [0,25],\quad
 n_s\in [0.9,1.1].
\end{equation}
For the cosmic shear data, we also allow one of the source redshift parameters to vary.
We choose $z_\mathrm{s}$, always marginalized in the final analysis, over its $2\sigma$ interval;
see Eq.~(\ref{eq:nzdeep05}) and~(\ref{eq:nzwide05}).

Indeed, one can criticize these assumptions but these are sufficient \emph{i)} to give us an
idea of the parameter space available for dark energy realized by quintessence and
\emph{ii)} to discuss to which extent weak lensing can improve the constraints on dark
energy models. From a pragmatic point of view, we were limited by computational capacities
and a more complete analysis will follow. Let us stress that such an analysis requires an
investigation of the potential degeneracies of the dark energy parameters with the standard
cosmological parameters and in particular, one would need to quantify how allowing for dark
energy changes the allowed range of variation of the other cosmological parameters. This is
left for further studies.

\begin{figure*}[thb]
 \centering
   \includegraphics[width=13cm]{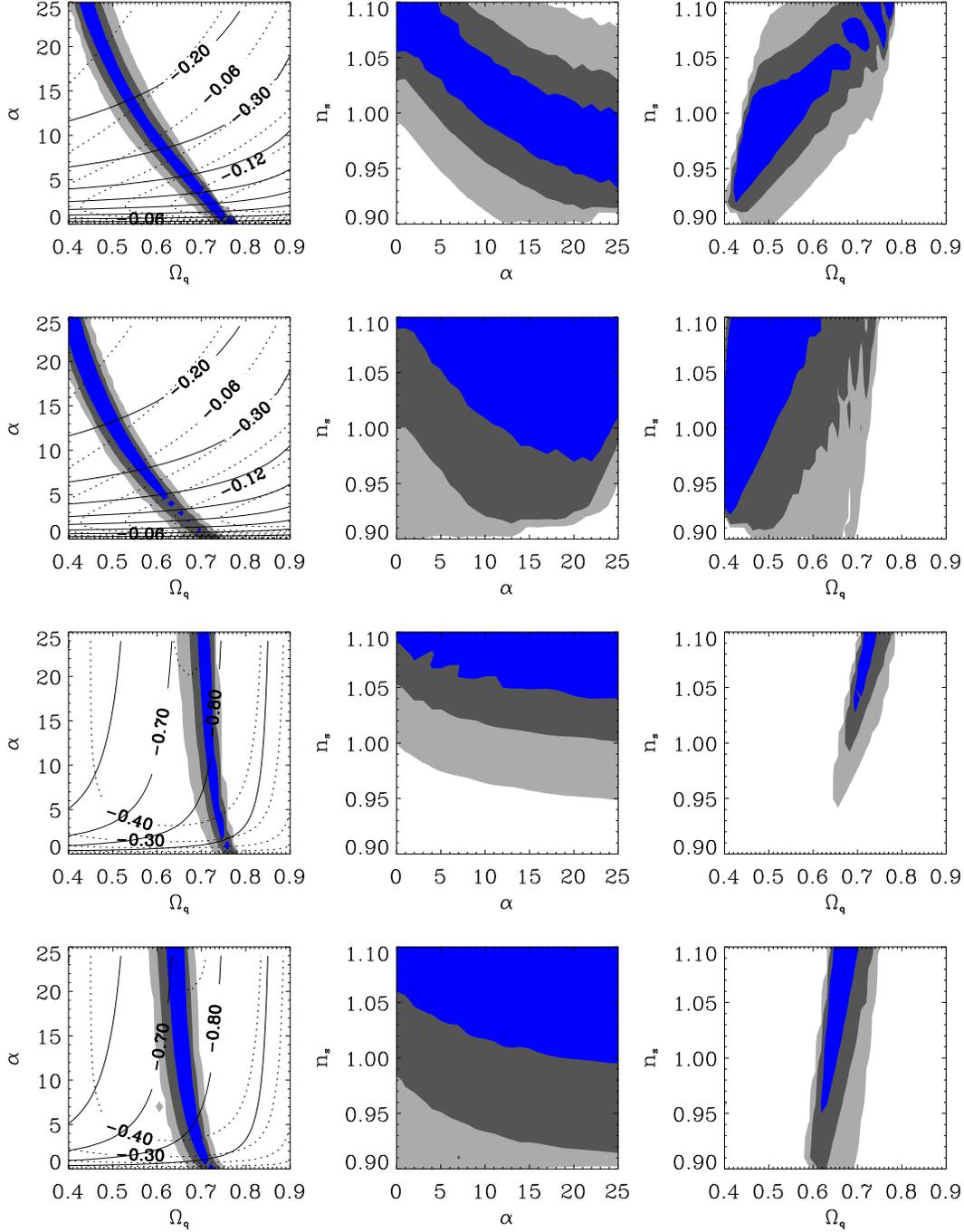}
 \caption{Joint likelihood analysis of VIRMOS-Descart, CFHTLS-deep, and
 CFHTLS-wide/22~deg$^2$ top-hat variance data -- contours at 68\%, 95\%, and 99\% C.L.
 for the variables $(\Omega_{\q 0},\alpha)$, $(\alpha,n_s)$ and $(\Omega_{\q 0},n_s)$.
 On the quintessence parameters sub-space, we have added the contour lines of
 $(w_\pivot,w_a)$ discussed in Fig.~\ref{fig:2b}, assuming $z_\pivot=0.5$. The
 two upper lines are dedicated to Ratra-Peebles models [Eq.~(\ref{eq:RP})] and
 the two lower to SUGRA models [Eq.~(\ref{eq:SUGRA})]. For each class of models,
 the non-linear spectrum has been computed using both the Peacock \& Dodds (1996)
 procedure, first and third lines, and the halo model approach by Smith~\etal (2003),
 second and fourth lines. See \S~\ref{sec3d} for discussion.} \label{fig:3a}
\end{figure*}


\subsection{Likelihood analysis: Joint cosmic shear data}\label{sec3d}

Using both real and synthetic cosmic shear data, we perform the likelihood analysis aiming
to investigate to which extent the constraints on cosmological parameters, and in particular
the quintessence ones, depend on the linear--to--non-linear mapping and on the selection
effects of the two-points statistics.

In this section, we focus on the first issue by combining top-hat variance data of
VIRMOS-Descart, CFHTLS-deep and CFHTLS-wide (22~deg$^2$ sub-sample) surveys; see
Table~\ref{tab:4} for numerical results of individual parameters constraints.
Figure~\ref{fig:3a} depicts the results quoting the 68\%, 95\% and 99\% confidence level
contours. In particular, we compute the non-linear spectra by both the Peacock \& Dodds
(1996) and Smith \etal (2003) procedures, for Ratra-Peebles and SUGRA models as well, hence
allowing for two kinds of comparisons.

Firstly, we can compare Ratra-Peebles and SUGRA likelihood contours, disregarding the
non-linear mappings. There are two striking differences. One is the strong constraint on
$\Omega_{\q 0}$ found in the SUGRA case. The reason for this is that, since the amplitude of
the power spectrum is kept fixed, the $(\sigma_8,\Omega_{\mat0})$ degeneracy implies a
strong constraint on $\Omega_{\mat0}$ and consequently on $\Omega_{\q 0}$ through the flat
universe prior. This effect is stronger in SUGRA models than in Ratra-Peebles ones since the
former approach a $\Lambda$CDM at low redshift. The other difference is the well defined
degeneracy found in the $(\Omega_{\q 0},\alpha)$ plane for the Ratra-Peebles case. This
feature will allow to put a stronger constraint on Ratra-Peebles' $\alpha$ than in SUGRA's
$\alpha$ when combining with other data.

Secondly and perhaps more interestingly, by comparing the non-linear mappings, for both
Ratra-Peebles and SUGRA models two classes of cosmological parameters come out: Those
concerning the quintessence, $\alpha$ and $\Omega_{\q 0}$, are found to be essentially
independent of the non-linear mapping used. This supports the claim of \cite{simpsonbridle}
that cosmic shear is sensitive to dark energy mostly through the background dynamics. The
second class involves the cosmological parameters accounting for the primordial universe,
here $n_s$ only. In this case the corresponding likelihood contours strongly depend on the
chosen mapping, regardless of the quintessence potential we used. Hence it seems not
possible to constrain the primordial spectral index, at least jointly with quintessence
parameters, until a stable formulation of the non-linear regime of structure formation will
be available.

\begin{figure*}[thb]
 \centering
   \includegraphics[width=12cm]{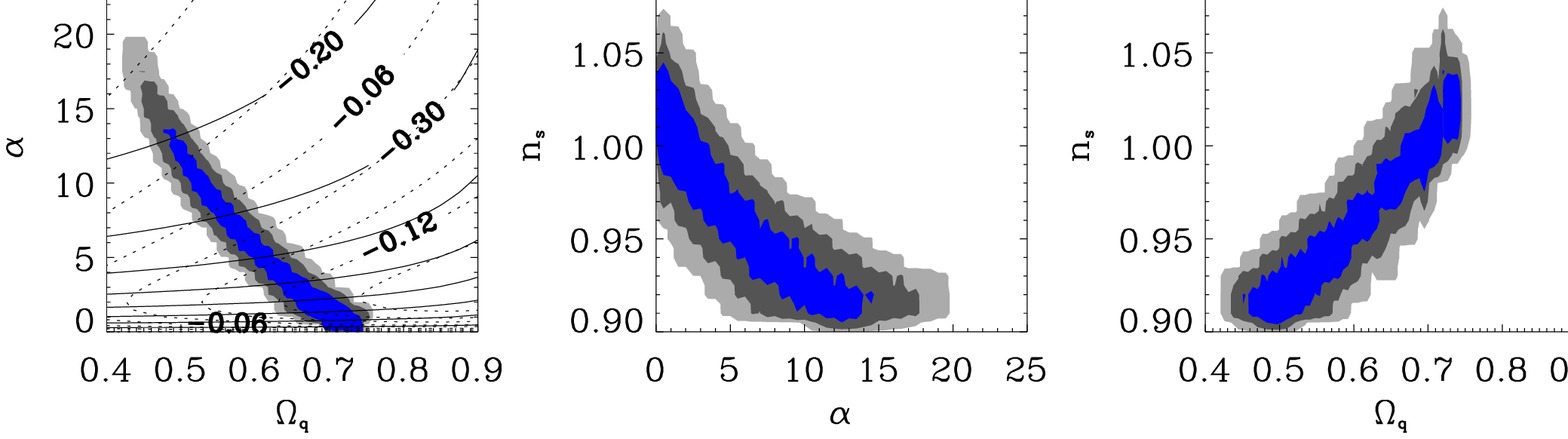}
 \caption{CFHTLS-wide constraints on Ratra-Peebles models.
 From left to right, we present the likelihood analysis (68, 95, and 99\% C.L.)
 for the variables $(\Omega_{\q 0},\alpha)$, $(\Omega_{\q 0},n_s)$ and
 $(\Omega_{\q 0},\alpha)$ and we have added the contour lines discussed in
 Fig.~\ref{fig:2b}, assuming $z_\pivot=0.5$. The first line depicts the
 analysis of the actual dataset based on 22~deg$^2$ (W1+W3; see \S~\ref{sec3e})
 while the three other lines are based on the synthetic data for a field
 of 170~deg$^2$. The second and third line show the top-hat shear variance
 and aperture mass variance, respectively. The fourth line describes the
 analysis of the same simulated data for the top-hat shear variance but
 using only angular scales larger than $20$~arcmin to cut out the non-linear
 part of the matter power spectrum. In particular, we conclude from the
 left column that the parameters describing the quintessence sector ($\alpha$
 and $\Omega_{\q 0}$) are not affected by the choice of the statistics and
 are well estimated by the linear part of the power spectrum. Here we use
 the Peacock \& Dodds (1996) mapping for the non-linear power spectrum. See
 \S~\ref{sec3e} for discussion.}
 \label{fig:3c}
\end{figure*}

\begin{figure*}[thb]
 \centering
   \includegraphics[width=13cm]{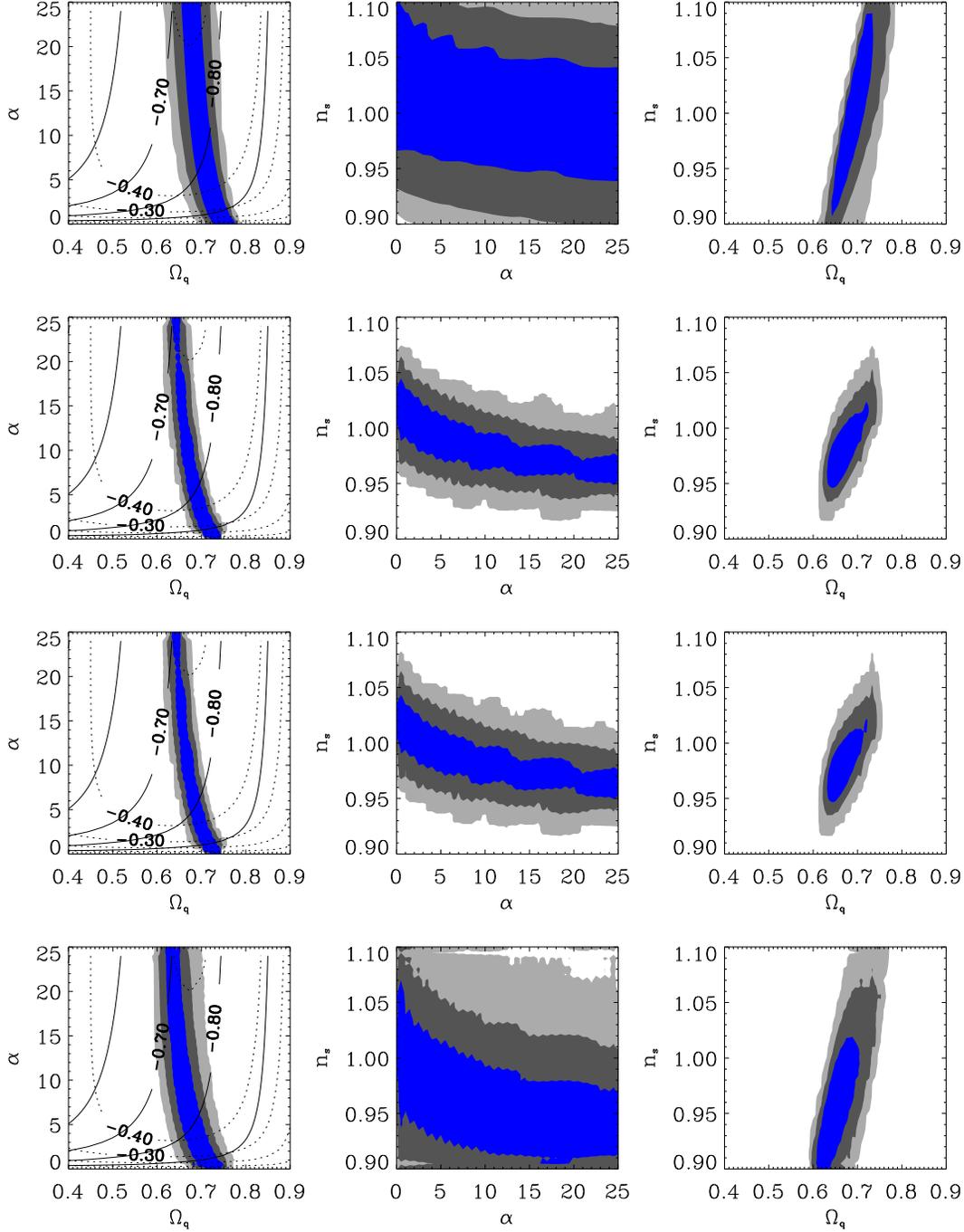}
 \caption{CFHTLS-wide constraints on SUGRA models. This figure is
 analogous to Fig.~\ref{fig:3c}. We reach the same conclusion as
 for Fig.~\ref{fig:3c}. We recover that the contours are almost
 independent on the value of $\alpha$. See \S~\ref{sec3e} for
 discussion.} \label{fig:3d}
\end{figure*}


\subsection{Likelihood analysis: Synthetic data}\label{sec3e}

\begin{figure*}[tbh]
 \centering
   \includegraphics[width=8.5cm]{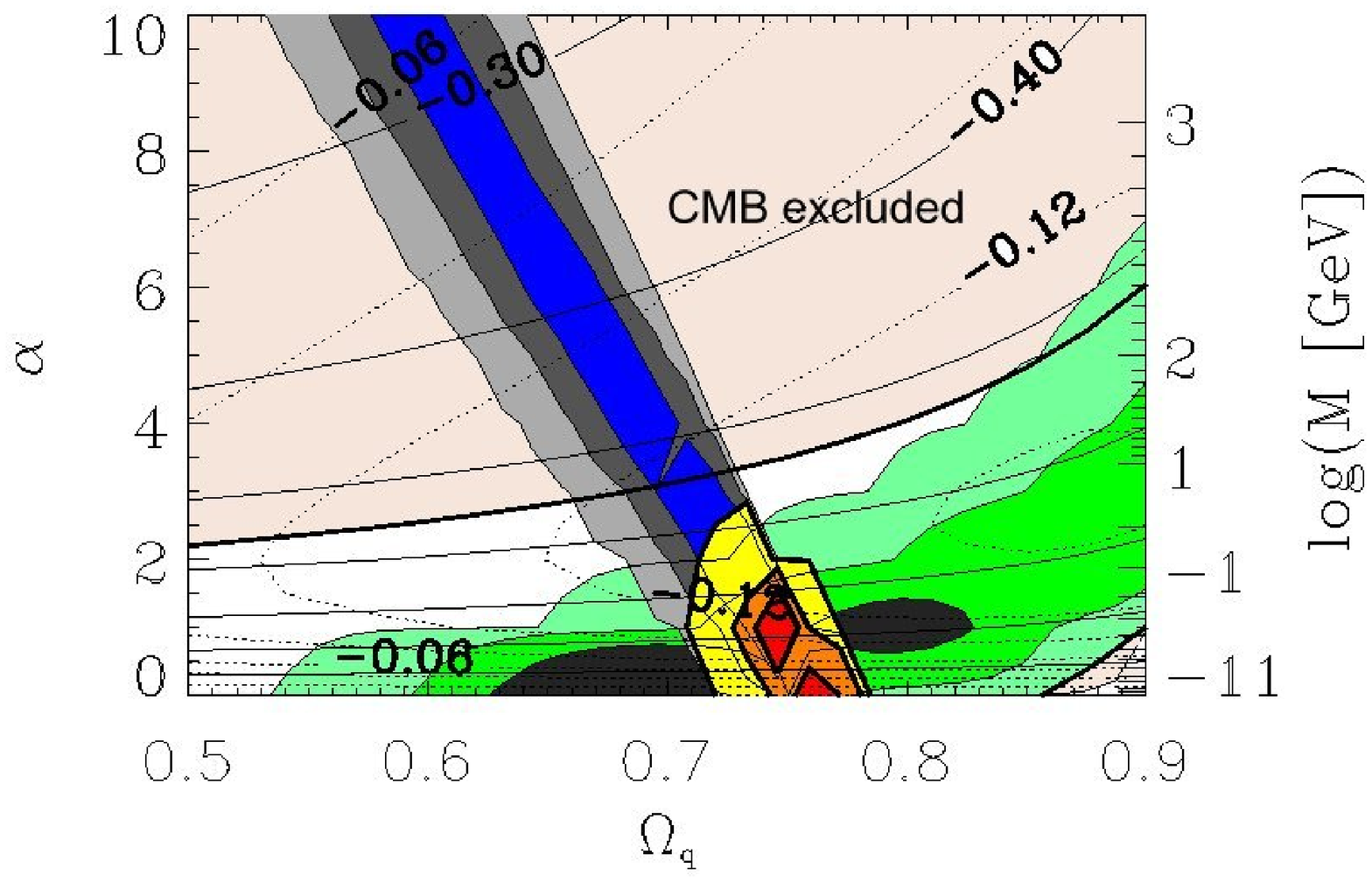}
  \hskip 0.5cm
   \includegraphics[width=8.5cm]{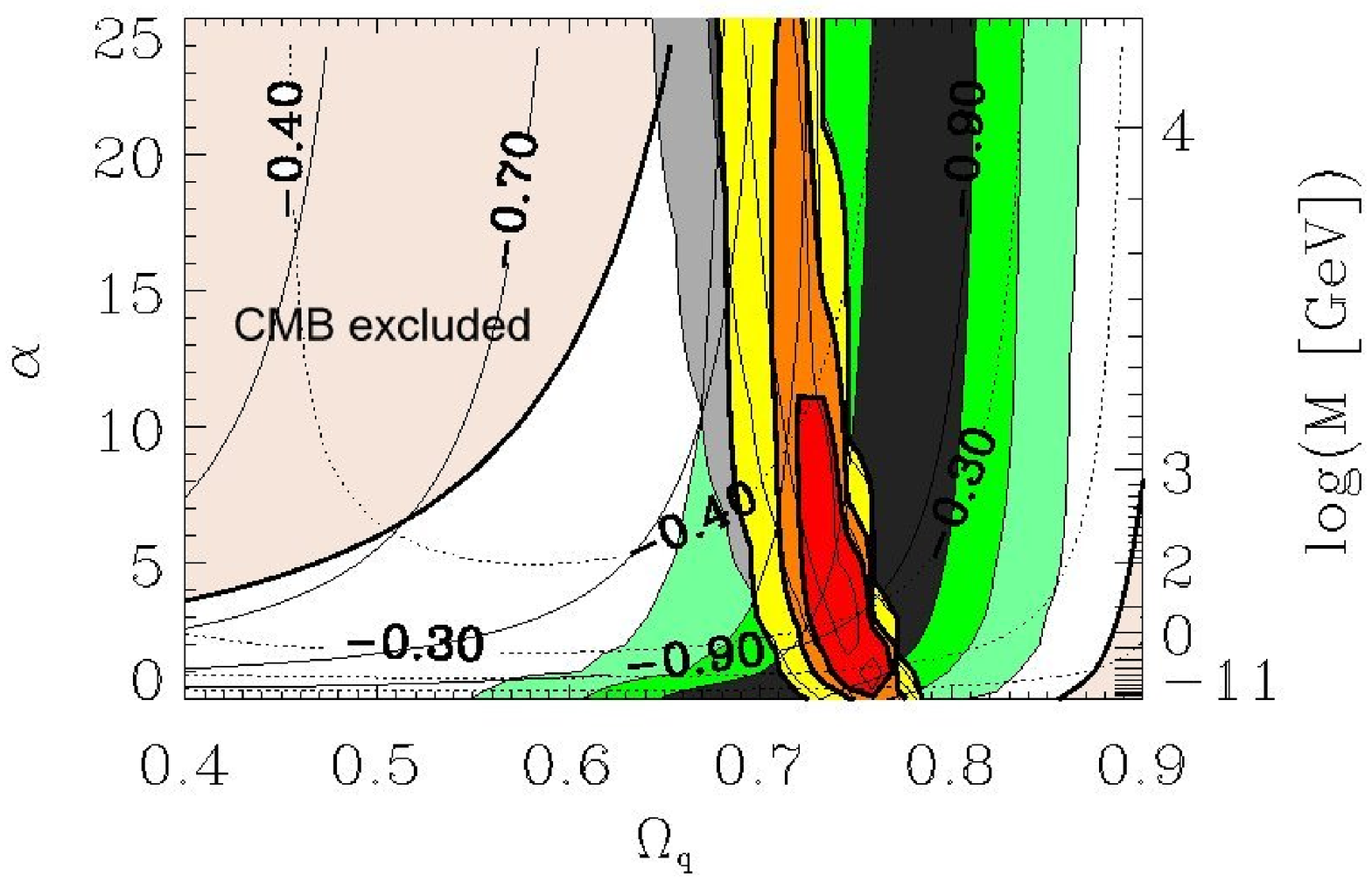}
  \vskip 0.7cm
 \caption{Joint analysis of the quintessence parameters $(\Omega_{\q 0},\alpha,)$ from CMB,
 Sn~Ia and cosmic shear (top-hat variance), for Ratra-Peebles models (left panel) and SUGRA
 models (right panel). The likelihood analysis uses the ``gold'' set for Sn~Ia (green contours),
 VIRMOS-Descart~+~CFTHLS-deep~+~CFTHLS-wide(22~deg$^2$) top-hat variance data for the cosmic
 shear (blue contours), and both combined (red contours). Contours correspond to 68, 95, and
 99\% C.L. According to WMAP-1yr measurements, the location of the first acoustic peak of the TT
 power spectrum of CMB, allowing for binning between neighboring multipoles, excludes regions of
 the parameter space (shadowed) nearly degenerate with Sn~Ia constraints; see \S~\ref{sec4b} for
 details. On the right axes, we quote an indicative mass scale of the quintessence potential,
 weakly dependent on $\Omega_{\q 0}$ but $\Omega_{\q 0}$ is small; see \S~\ref{sec2} for details.
 Finally, contours of $w_\pivot$ (solid) and $w_a$ (dotted) are superposed, setting $z_\pivot=0$.
 As for Ratra-Peebles models, supernovae data put strong constraints on $\alpha$ which are tighten
 by cosmic shear, while for SUGRA models all observables are fairly insensitive to the value of
 this parameter. The corresponding constraints are given on Table~\ref{tab:4}.}
 \label{fig:4a}
\end{figure*}

Using wide surveys, we can eventually investigate the effect of the non-linear regime of structures
formation. For this purpose, we perform a likelihood analysis using a synthetic realization of
the full CFHTLS-wide survey. This consists on synthetic data vectors of top-hat and aperture
mass variance and on a synthetic covariance matrix. The former are computed at a fiducial model,
which we take to be a $\Lambda$CDM with the current CFHTLS-wide redshift distribution. The
covariance matrix is computed using the analytical approximation derived in \cite{schneider02}.
It depends on three main features of the survey; the effective area $A$, the effective galaxy
number density $n_{\rm gal}$ and the dispersion of the distribution of ellipticities $\sigma_e$.
For these parameters we used the values that are expected at the end of the CFHTLS-wide campaign,
respectively: $A=170$~deg$^2$, $n_\mathrm{gal}=20$~gal/arcmin$^2$, and $\sigma_e=0.4$. Notice we
assume a larger ellipticity dispersion and a higher density of galaxies than those obtained in
Hoekstra \etal (2005). They correspond to a different galaxy weighting scheme than the one used
with current data.

Figures~\ref{fig:3c} and~\ref{fig:3d} outline the likelihood analysis of Ratra-Peebles and
SUGRA models, respectively -- contours at 68\%, 95\% and 99\% confidence level. We show only
the results achieved using the mapping by Peacock \& Dodds (1996), those achieved using the mapping
by Smith \etal (2003) being in agreement at a 10\% level.

Firstly, by comparing the results achieved using the top-hat variance data of the 22~deg$^2$
sub-sample (first line) with those of the synthetic $170$~deg$^2$ field (second line), it is
evident the gain achievable by the full survey. In particular, it is worth noticing that the
pure quintessence parameters sub-space is less dependent on the survey area, while the
constraints on the primordial spectral index strongly depend on it. Hence, the distinction
of cosmological parameters in two classes introduced in the previous section seems
confirmed.

The second and third lines show the likelihood contours for the top-hat and aperture mass
variances, respectively. They are consistent, depicting compatible confidence level regions,
for both Ratra-Peebles and SUGRA models. There are two main differences in the properties of
aperture mass and top-hat variances. Measurements of aperture mass variance at different scales
are less correlated than top-hat variance ones, since the former is a narrow filtered version
of the shear power spectrum; and their effective range extend to only about $1/5$ of the top-hat
variance one. Hence they follow more accurately the shape of the power spectrum at their measured
range. For a power spectrum featureless at most scales but these, the independent measurements
render this statistic the most convenient to use. In this case, contours from aperture mass
variance are expected to be smaller than top-hat variance ones. But in general, the loss of
information regarding the behaviour of models at other scales diminishes the capability of
distinguishing between models. Consequently, the aperture mass variance contours are, in
general, expected to be larger than top-hat variance ones. The difference is larger when
analysing cosmic shear most important parameters. Since we are studying parameters to which
cosmic shear is moderately sensitive to, the difference is not noticeable in our results of
Fig.~\ref{fig:3c} and~\ref{fig:3d}, but it is striking when using VIRMOS-Descart data to
constrain the $(\sigma_8,\Omega_m)$ plane (\cite{vw01}). Furthermore, for our data with
measurements up to less than 1~deg, the largest power spectrum scale probed by the aperture
mass variance is the one probed by the top-hat variance at around 10~arcmin and its full
range effectively lies on non-linear scales, rendering its infered parameters' constraints
less reliable, due to non-linear modeling uncertainties. For these two reasons, we will
choose to use top-hat variance only, when producing the final results.

By using the full wide survey, we can try to disentangle the effects of the
non-linear regime of structures formation by cutting off the small angular scales from the
final analysis. In such a way, we can better investigate the distinction of cosmological
parameters in the two classes discussed above, probing if dark energy primarly hangs on the
background dynamics. The plots on the bottom line of Fig.~\ref{fig:3c} and~\ref{fig:3d}
depict the analysis of the top-hat variance when taking into account only angular scales
larger than 20~arcmin, corresponding to wavevectors $k\la 1 h$~Mpc$^{-1}$ at $z\la\langle
z\rangle /2 \sim 0.5$, where therefore the effects of the non-linear regime are
sub-dominant. The spread in the likelihood contours is more relevant for plots involving the
primordial spectral index $n_s$, while the quintessence parameters $(\alpha,\Omega_{\q 0})$
are not so much affected. Hence, one can study dark energy by cosmic shear using this
technique even if not properly knowing how to deal with the non-linear regime. Obviously
this conclusion has to be confirmed by a more complete study involving a larger parameter
space, to account for other degeneracies.

\begin{table*}
\caption{Results of separate and joint analysis of cosmic-shear and Sn~Ia data on
 quintessence parameters $(\alpha,\Omega_{\q 0})$ at 68\% (95\%) confidence level.
 For Ratra-Peebles models, we quote only upper limits for $\alpha$, the best-fit being
 always at $\alpha=0$. Only for indicative purposes, we quote 68\% confidence level limits
 on $(w_\pivot,w_a)$ parameters computed at $z_\pivot=0$, setting \emph{n.c.} when not
 constrained. Remind that in this class of models $w_0\geq-1$. See \S~\ref{sec4a} for
 discussion.}
\label{tab:4}
\centering
\begin{tabular}{l c c c c|c c c c}
\hline\hline
   & \multicolumn{4}{c}{Ratra-Peebles} & \multicolumn{4}{c}{SUGRA}  \\
   & $\alpha$                      & $\Omega_{\q 0}$ & $w_0$ & $w_a$ & $\alpha$ &
   $\Omega_{\q 0}$ & $w_0$         & $w_a$ \\
\hline \\
  $\langle\gamma^2\rangle$         & $<14\,(25)$ & $0.63^{+0.14\,(+0.15)}_{-0.17\,(-0.20)}$ & n.c.                         & n.c.                      & $13^{+12\,(+12)}_{-5\,(-13)}$ & $0.72^{+0.01\,(+0.03)}_{-0.02\,(-0.05)}$ & $\la -0.79$ & $\ga -0.45$ \\
  Sn~Ia                            & $<1\,(3)$   & $0.74^{+0.09\,(+0.16)}_{-0.10\,(-0.11)}$ & $\la -0.75$    & $\ga -0.13$ & $12^{+5\,(+11)}_{-12\,(-12)}$ & $0.77^{+0.04\,(+0.07)}_{-0.04\,(-0.08)}$ & $\la -0.84$               & $\ga -0.43$ \\
  $\langle\gamma^2\rangle$~+~Sn~Ia & $<1\,(1)$   & $0.75^{+0.02\,(+0.03)}_{-0.04\,(-0.04)}$ & $\la -0.7$     & $\ga -0.13$ & $2^{+7\,(+18)}_{-1\,(-2)}$    & $0.74^{+0.03\,(+0.03)}_{-0.04\,(-0.05)}$ & $\la -0.84$               & $\ga -0.38$ \\
\hline
\end{tabular}
\end{table*}

The first line of Fig.~\ref{fig:3c} and~\ref{fig:3d} show the current results of CFHTLS-wide,
using W1 and W3 data from an effective sky coverage of 22~deg$^2$. The corresponding marginalized
constraints on $\alpha$ and $\Omega_{\q 0}$ are a factor of 2 larger than the ones found with the
synthetic data, depicted on the second line of the same figures. This is consistent with the
reduction of the data error bars, which are proportional to $\sigma_e^2\,n_\mathrm{gal}^{-1}\,
A^{-1/2}$, if cosmic variance is not taken into account. We must caution that the gain in the
parameters space cannot be estimated with precision by this simple argument, namely by assuming
\begin{equation}\label{gain}
{\rm gain} \sim \frac{r^2_{\sigma_e}}{r_{n_{\rm gal}}\,\sqrt{r_A}},
\end{equation}
where the several factors $r$ are the ratios between the features of the two surveys. In fact,
the gain in the data error bars does not translate linearly into a gain in the parameters space
confidence levels. That happens only in the Fisher matrix approximation, and even there, only in
the case of statistical uncorrelated parameters. Furthermore, this reasoning does not take into
account the extra constraining power coming from measurements at larger scales, as discussed
earlier on, or simply coming from the fact of disposing of more degrees of freedom for the
likelihood calculations.

It is also worthwhile to notice that, in all cases, (from Fig.~\ref{fig:3a} to Fig.~\ref{fig:3d}),
a $\Lambda$CDM model with a Harrison-Zel'dovich spectrum [($\alpha,n_s)=(0,1)$] is compatible with
the data for $\Omega_{\q 0}\sim0.7$ at 99\% confidence level.


\section{Combining with other observables}\label{sec4}


\subsection{Sn~Ia}\label{sec4a}

We combine the cosmic shear data by VIRMOS-Descart and CFHTLS-deep and -wide (22~deg$^2$
sub-sample) surveys with the type~Ia supernovae  ``gold'' set by Tonry \etal (2003). In
particular, we evaluate confidence intervals considering cosmic shear and Sn~Ia both separately
and jointly. Indeed, since the distance modulus depends only on the background dynamics, we can
restrict to the quintessence cosmological parameters $\alpha$ and $\Omega_{\q 0}$.

Figure~\ref{fig:4a} depicts both the independent and combined analysis for Ratra-Peebles and
SUGRA models, using the top-hat variance data for the cosmic shear and computing the
non-linear spectrum by the Peacock \& Dodds (1996) mapping. The corresponding results are
summarized in Table~\ref{tab:4}; these marginalized results for each parameter do not assume
any prior knowledge of the other, apart from the flat priors implied by the range of the
grid. Let us emphasize that the constraints obtained on $(w_0,w_a)$ inferred from those on
the parameters of the potential differ when we change the form of the potential. This
confirms that constraints on the equation of state derived from a general parameterization
have to be interpreted with care.

Concerning Ratra-Peebles models, the weak lensing and Sn~Ia contours are closer to mutual
orthogonality, so narrow joint constraints are expected. Data strongly favor a quintessence
component close to a cosmological constant, the best-fit lying always at $\alpha=0$, hence
in Table~\ref{tab:4} we prefer to present the constraints as an upper limit. As it is clear
from Fig.~\ref{fig:4a}, Sn~Ia are much more constraining for this parameter than weak
lensing. However, even though Sn~Ia alone reject $\alpha\geq 1$ at the 68\% level, as was
well known (\cite{SnIaQ}), the information of weak lensing further narrows the interval.

As far as SUGRA models are concerned, for a wide interval of $\alpha$ (approximately
$\alpha>5$) both the luminosity distance and the shear two-point correlations are (almost)
independent of this parameter, as it is clear from Fig.~\ref{fig:4a}. In fact, this
conclusion was already reached studying the CMB temperature anisotropies (\cite{braxetal}),
arguing that the equation of state does not strongly depends on the slope of the potential
leaving both distances and linear growth factor almost unchanged. In terms of statistical
significance, this means that the likelihood with respect to Sn~Ia or weak lensing data
alone is almost flat and so the best fit value has little meaning (see also \cite{SnIaQEQ}).
In Table~\ref{tab:4} we list the results in the form of a parameter value, which we take to be
the likelihood weighted average of $\alpha$, plus or minus the necessary deltas to form the
confidence intervals obtained. However, for a substantial interval of $\alpha$, weak lensing
and Sn~Ia lead to different, almost non-intersecting, ranges of $\Omega_{\q 0}$: Sn~Ia favoring
a higher value of $\Omega_{\q 0}$ while weak lensing a lower one. Thus, even though both
observables have limited sensitivity to constrain the parameter $\alpha$, this fact allows
to obtain a reasonable constraint from the joint likelihood, as that interval of $\alpha$ is
rejected by the data. The joint confidence interval are substantially reduced, with a
distinctive maximum of likelihood at $\alpha=2$. Notice, however, that this is an unstable
situation. In fact, if cosmic shear or supernovae contours slightly change their orientation and
size, due for example to a larger uncertainty on the redshift of sources, the joint contour will
easily degradate. On the contrary, the extreme situation, with both contours vertical and parallel
for all $\alpha$, would imply the abandon of the subjacent quintessence model. It has to be
stressed that every systematic effect on weak-lensing or supernovae data as well, relying for
instance on the data analysis procedure, strongly affects the final result; therefore special
care is necessary when combining several datasets.

Provided that it is not evident which pivot redshift should be used when parameterizing these
quintessence models by an equation of state of the form of Eq.~(\ref{eq:genLinder}), we superpose
to the $(\Omega_{\q 0},\alpha)$ plane, in Fig.~(\ref{fig:4a}), the contours for $w_\pivot$ and
$w_a$ corresponding to $z_\pivot=0$. Let us stress that the final results in terms of the equation
of state parameters are only indicative. Interestingly, notice from Fig.~\ref{fig:2b}
that for Ratra-Peebles models the estimation of $w_\pivot$ and $w_a$ should not change if
using $z_\pivot=0.5$ instead of $z_\pivot=0$. In fact, for this class of models, noticeably
different values on these parameters only appear when considering a high pivot redshift. On the
contrary, for SUGRA models the choice of the pivot redshift would be relevant already at low
redshift. See Table~\ref{tab:4} for specific constraints on $w_0$ and $w_a$.

\begin{figure*}[thb]
 \centering
   \includegraphics[width=7cm]{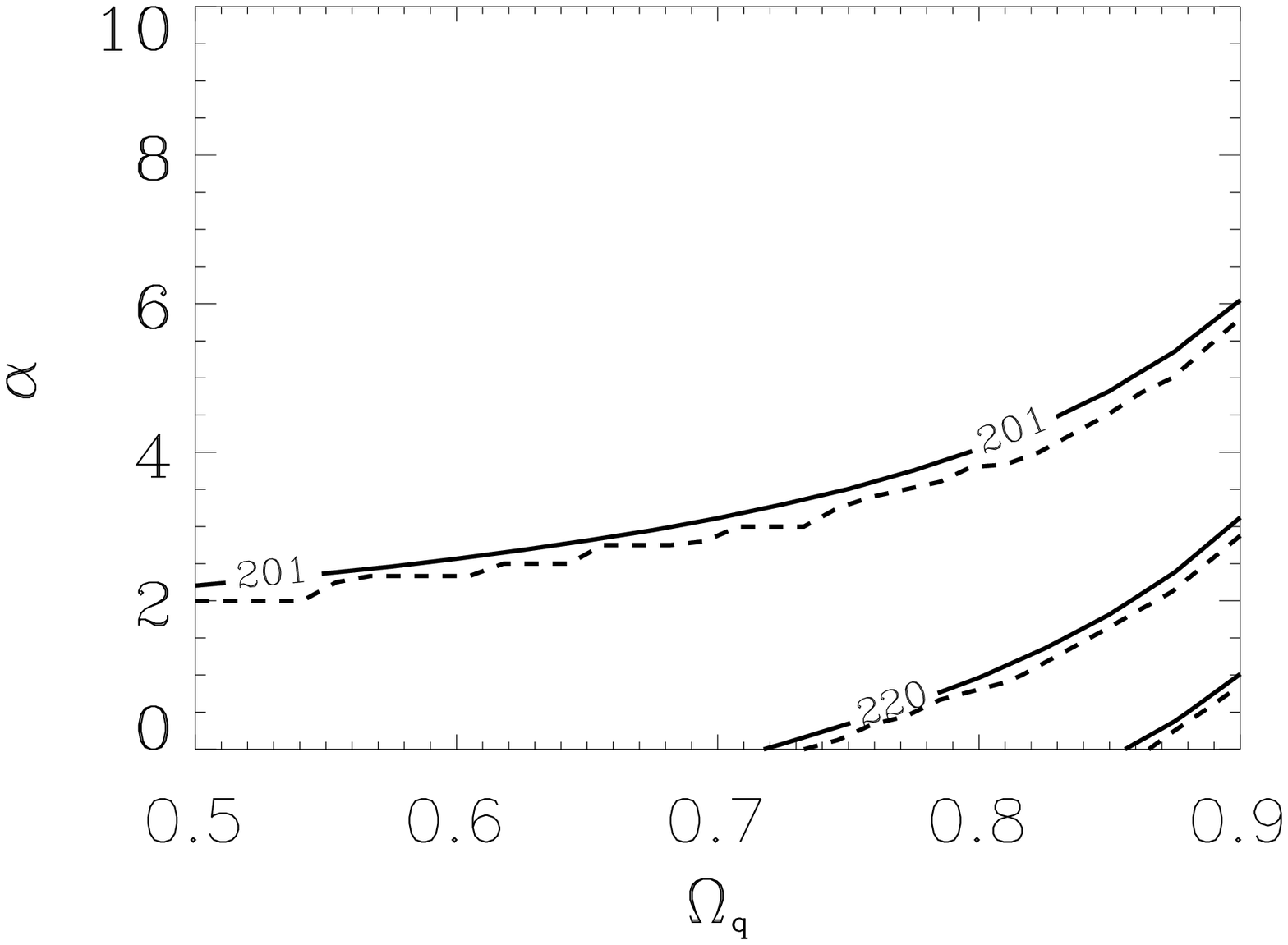}
\hskip 0.5cm
   \includegraphics[width=7cm]{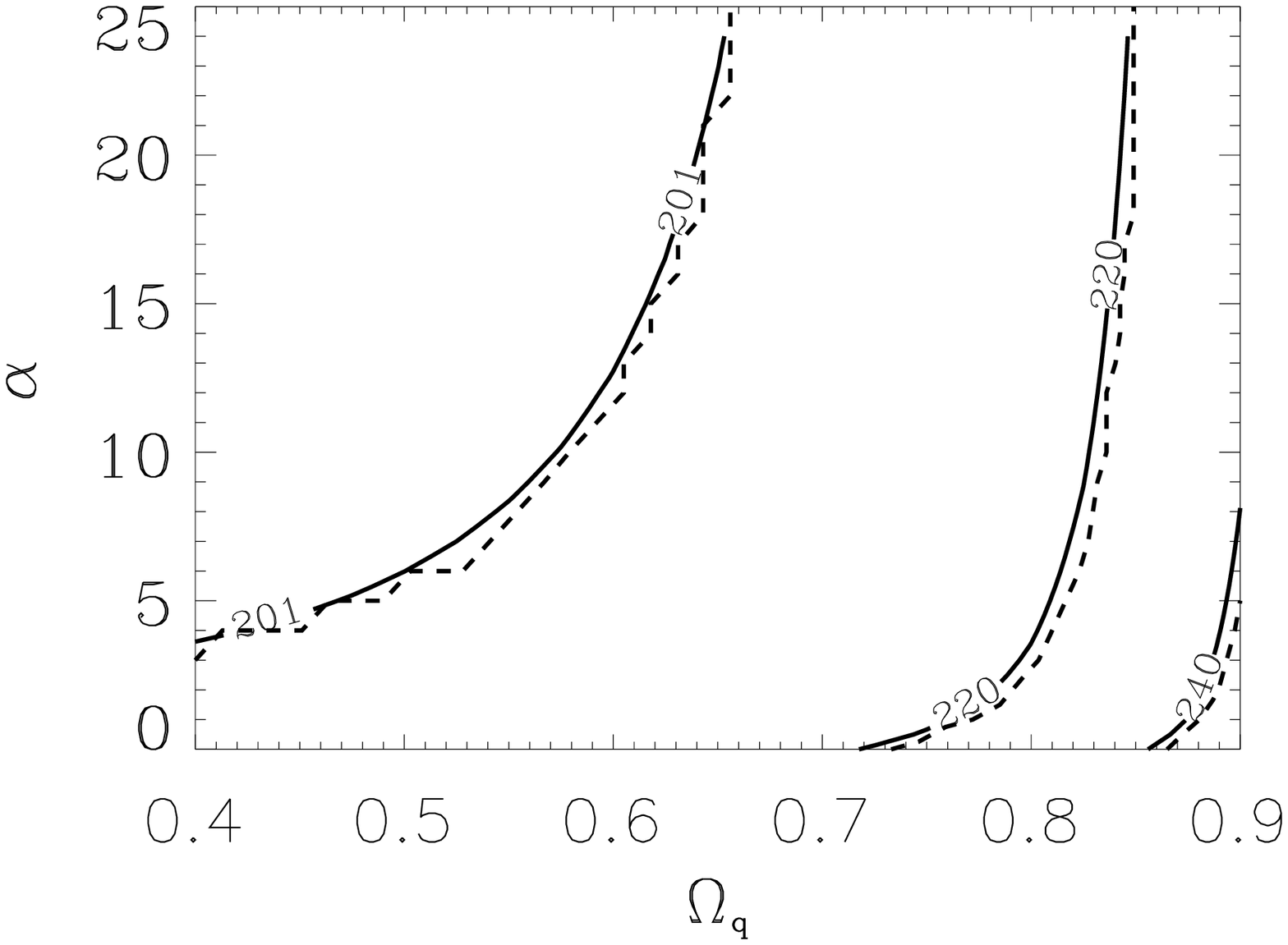}
\vskip 0.5cm
   \includegraphics[width=7cm]{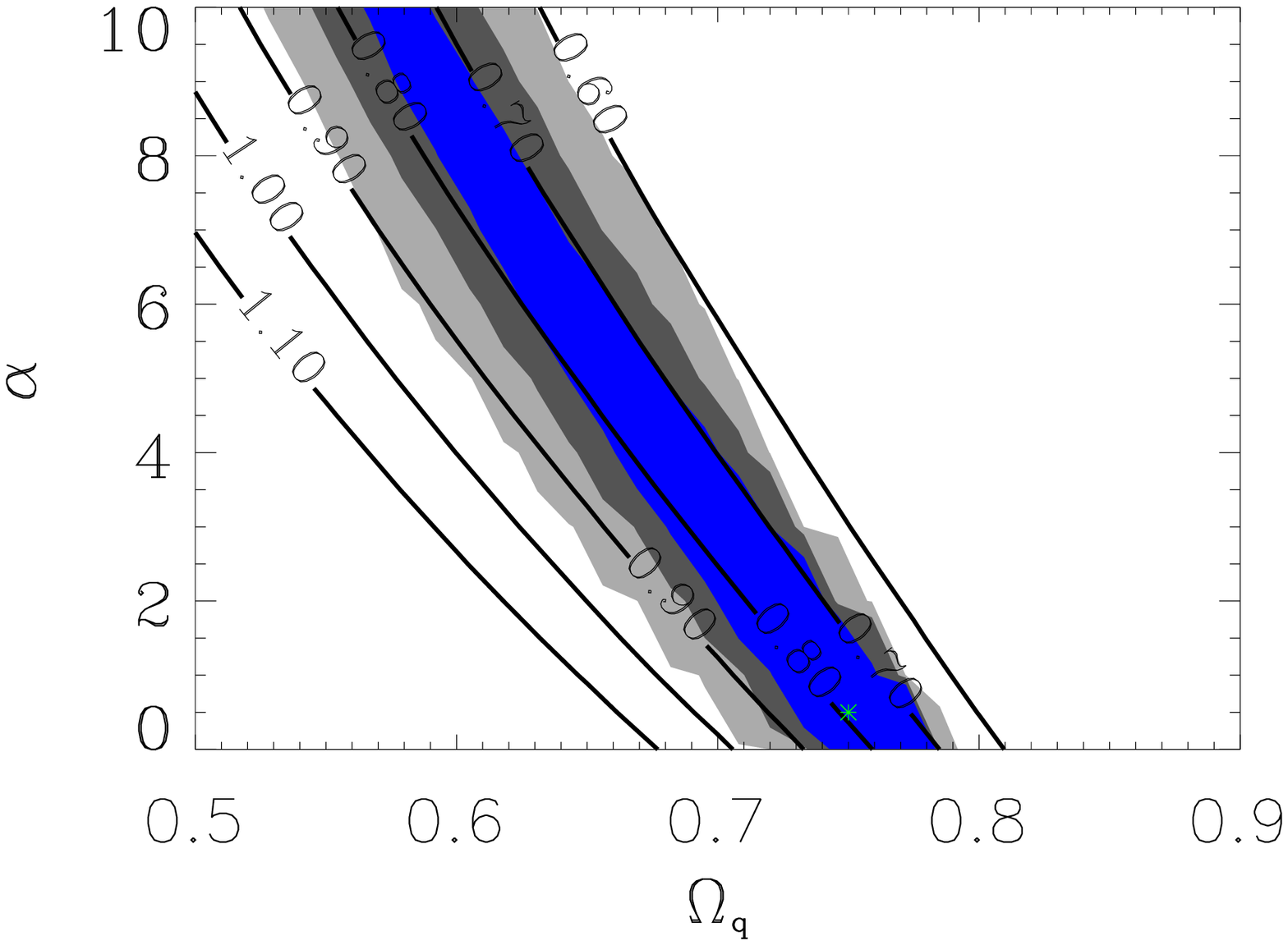}
\hskip 0.5cm
   \includegraphics[width=7cm]{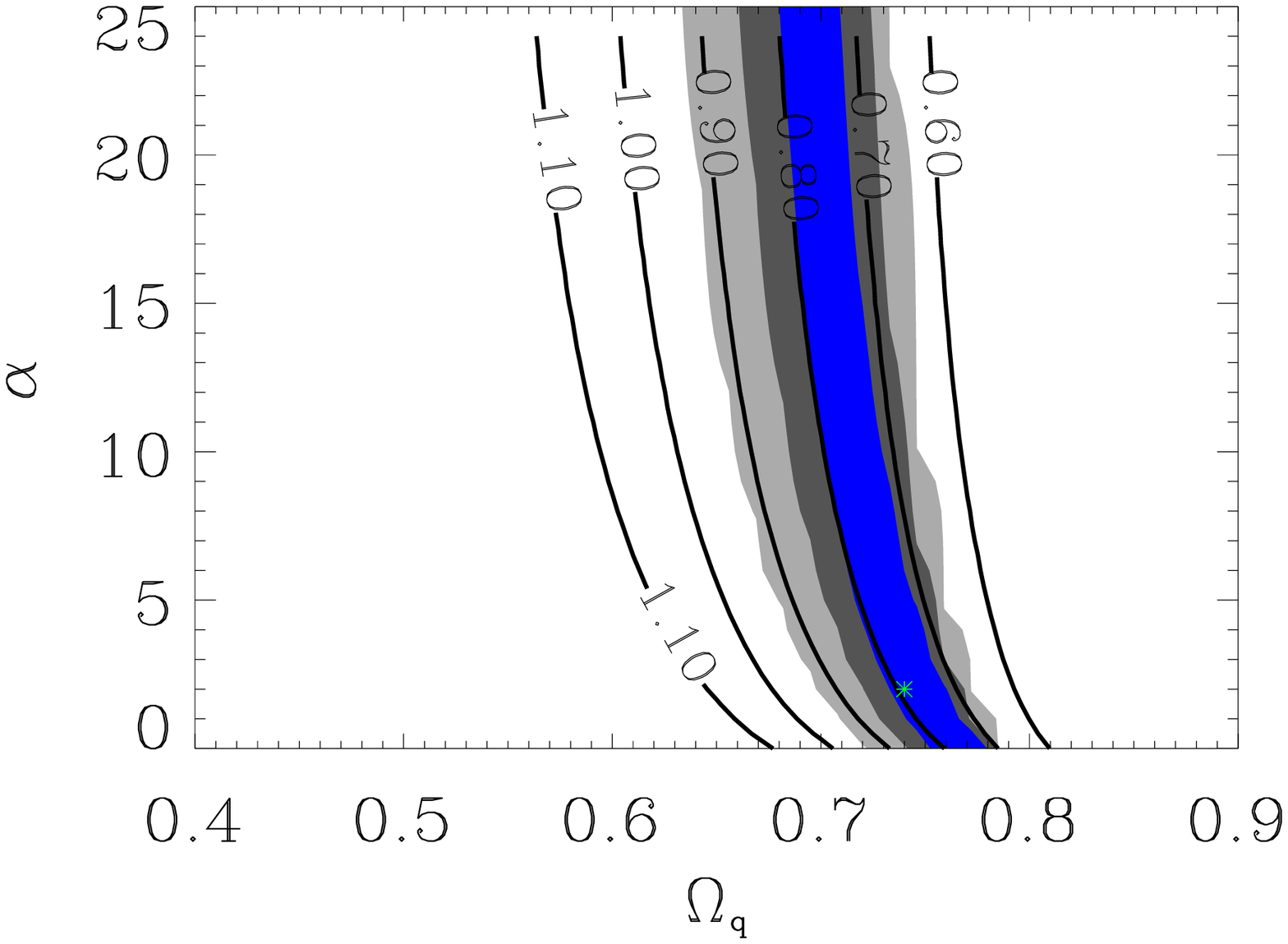}
\vskip 0.7cm
 \caption{Top line: Location of the first acoustic peak of the temperature CMB
 power spectrum, for Ratra-Peebles (left panel) and SUGRA (right panel) models.
 It is estimated using an analytical approximation lying on the solution
 of the background equations only (solid line), or by the full computation
 of $C_\ell$~s (dashed line). The central set of lines corresponds to the
 location of the first peak according to the best-fit  of WMAP-1yr data,
 $\ell=220.1\pm0.8$ (\cite{WMAP3}), while the left and right ones correspond
 to the  multipoles $\ell=201$ and $\ell=240$, defining the smallest and
 the largest multipoles contributing to the three points of binned data
 defining the first peak (\cite{WMAP}).
 Bottom line: Contour levels of $\sigma_8$ for Ratra-Peebles (left panel) and SUGRA
 (right panel) models with $n_s=1$, superposed to the cosmic shear data contours (VIRMOS-Descart
 + CFHTLS-deep + CFHTLS-wide/22deg$^2$). For indicative purposes, the star point
 marks the best-fit from the joint cosmic shear and SN~Ia analysis, roughly sitting at
 $\sigma_8=0.8$. Remind that the spectrum was normalized on the CMB so this plots show
 the consistency with a normalization at $z=0$. See \S~\ref{sec4b} for details.}
 \label{fig:isol}
\end{figure*}

Recently, weak lensing data (\cite{CFHTLSdeep05}; \cite{CFHTLSwide05}) were used to put
constraints on $w$ under the assumption it is a constant parameter. In that particular case,
one has two characteristic redshifts, $z_a$ and $z_\de$, defining the beginning of the
acceleration phase [$\ddot a(z_a)=0$], and of the domination of the dark energy [$\Omega_
{\mat}(z_\de)=\Omega_\de(z_\de)$]. They are given by
\begin{equation}
 (1+z_a)^{3w}=-\frac{1}{1+3w}\frac{\Omega_{\de 0}}{\Omega_{\mat 0}},\qquad
 (1+z_\de)^{3w}=\frac{\Omega_{\de 0}}{\Omega_{\mat 0}}
\end{equation}
so that $z_a>z_\de$. When $w$ becomes negative and large in absolute value, $z_a$ and $z_\de$
tend to zero so that dark energy just starts almost today to dominate the universe and is
redshifted in a way that it does not affect even low redshift observables. It follows that
we expect the data to be insensitive to the value of $w$ in that regime so that one can get
only an upper limit on its value, as found by Semboloni \etal (2005) and Hoekstra \etal (2005).
Indeed, such a situation cannot be achieved with a physical model as considered here because by
construction it imposes that $w>-1$. It follows that our approach is a physically motivated way
of imposing a prior on $w$. Note also that Fig.~\ref{fig:2} shows that $w=$~constant is not a
good approximation of these two classes of models. At best the constraint on a constant $w$ can
be related to some redshift average of the equation of state. For these reasons, it is difficult
to deduce a constraint on the physical models from a constraint on a constant $w$, even though
the results of Table~\ref{tab:4} are compatible with them.


\subsection{CMB}\label{sec4b}

CMB temperature anisotropies have been extensively used (see e.g.~\cite{baccigalupi02};
\cite{jassal}; \cite{corcmb}) but several
degeneracies amongst the cosmological parameters prevent to accurately constrain the
cosmological parameters using CMB data only. Hence, concerning weak lensing, several studies
already attempt to combine CMB and cosmic shear in order to constrain the cosmological
parameters (see e.g.~Contaldi, Hoekstra \& Lewis 2003; Ishak \etal 2004; Tereno \etal 2004).

\begin{figure*}[htb]
 \centering
   \includegraphics[width=8cm]{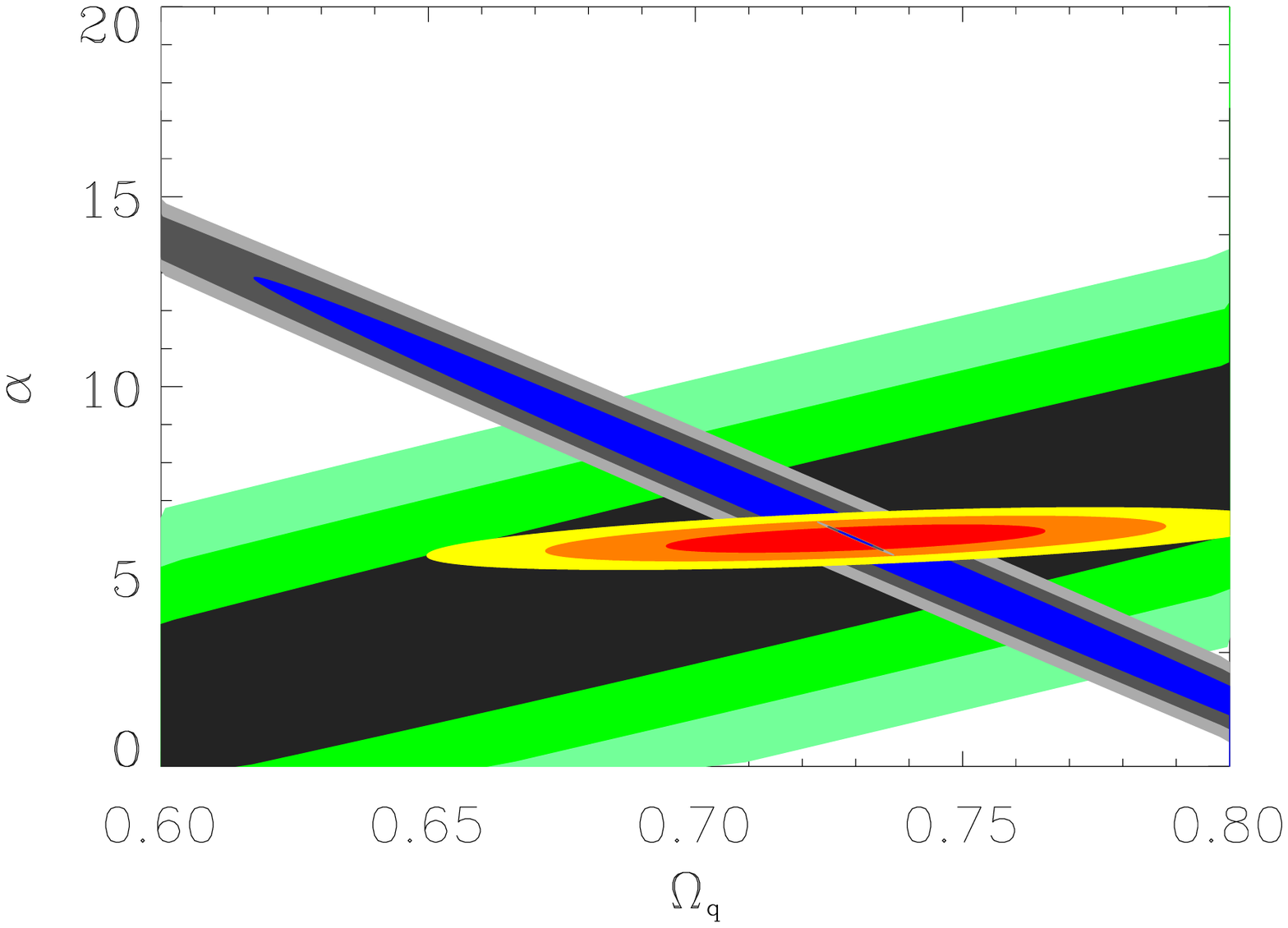}
   \includegraphics[width=8cm]{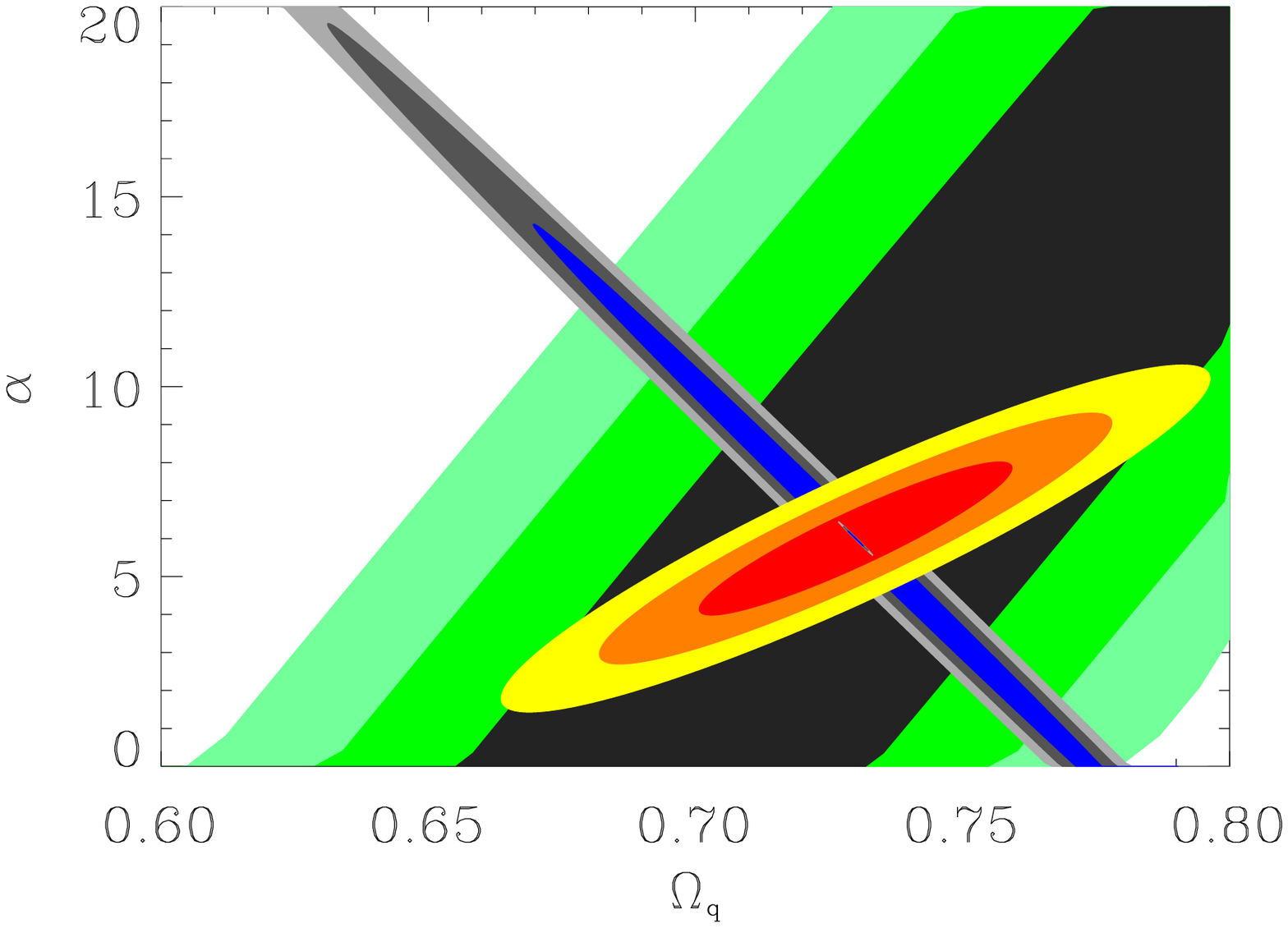}
 \caption{Fisher analysis for Ratra-Peebles (left) and SUGRA (right)
 models, for cosmic shear top-hat variance (blue ellipses), Sn~Ia
 ``gold'' set (green) and temperature CMB angular power spectrum by
 WMAP-1yr data (red) -- contours at 68\%, 95\%, and 99\% C.L. For the
 CMB, the noise matrix was computed using the public  code by \cite{WMAPverde}.
 For cosmic shear, we consider both CFHTL-wide survey ($A=170$~deg$^2$,
 $n_\mathrm{gal}=20$~gal/arcmin$^2$, $\sigma_e=0.4$) and a space based
 survey ($A=20,000$~deg$^2$, $n_\mathrm{gal}=35$~gal/arcmin$^2$, $\sigma_e=0.3$),
 quoted as type~II in the text. We consider only the quintessence
 parameters while the other parameters are kept {\it fixed}.
 See \S~\ref{sec5} for discussion.}
 \label{fig:5b}
\end{figure*}

As for dark energy, it affects the CMB anisotropies angular power spectrum at least in two ways
(\cite{braxetal}). Firstly, the angular diameter distance is modified so that the peak structure
is shifted. In particular the location of the first acoustic peak, depending on the geometry of
the universe, provides an estimate of the angular diameter distance to the last scattering surface.
However, also pre-recombination effects can shift the peaks from their true geometrical locations
(\cite{doran}; \cite{kamion}). Secondly, the time evolution of the dark energy strongly affects
the integrated Sachs-Wolfe effect. This effect is more relevant at low multipoles, modifying the
amplitude of the spectrum, but it also leads to an additional shift of the Doppler peaks.

Beside using CMB data to normalize the power spectra (see \S~\ref{sec3c}), we use them in still
another way, by noticing that even without a statistical analysis of CMB data, we can strongly
constrain cosmological parameters, and notably quintessence ones, simply by using the location
of the first Doppler peak. Moreover, even without solving the perturbations equations and computing
the TT spectrum, one can compute the acoustic scale just by solving the equation for the
background evolution. The location of acoustic peaks is then estimated allowing for the shifts
induced by the dark energy by means of a fitting formulae (\cite{doranlilley}). In conclusion,
the location of the first acoustic peak will be a function of the quintessence parameters
$(\Omega_{\q0},\alpha)$, with a negligible dependence on the primordial spectral index $n_s$ due
to the shift correction. Using this analytic approximation, we individuate, in Fig.~\ref{fig:4a},
the region of the $(\Omega_{\q 0},\alpha)$ plane compatible with the location of the first acoustic
peak of WMAP-1yr data (\cite{WMAP}; \cite{WMAP3}) including the bins' contributions to the three
points defining the peak, namely $201<\ell<240$. Consistently with the normalization procedure
we used, this result is in excellent agreement with that deduced by the complete computation of
the TT spectrum, as shown in Fig.~\ref{fig:isol} (top panels). Let us notice that the region of the
parameter space $(\Omega_{\q 0},\alpha)$ compatible with the position of the first acoustic peak is
degenerate with the Sn~Ia constraints; the reason being that, since the pre-recombination effects
of quintessence on the definition of the acoustic horizon at last scattering are negligible, the
acoustic scale eventually depends only on on the Hubble parameter like the luminosity distance,
hence both are ultimately affected by quintessence approximatively in the same way.

\begin{figure*}[htb]
 \centering
   \includegraphics[width=17cm]{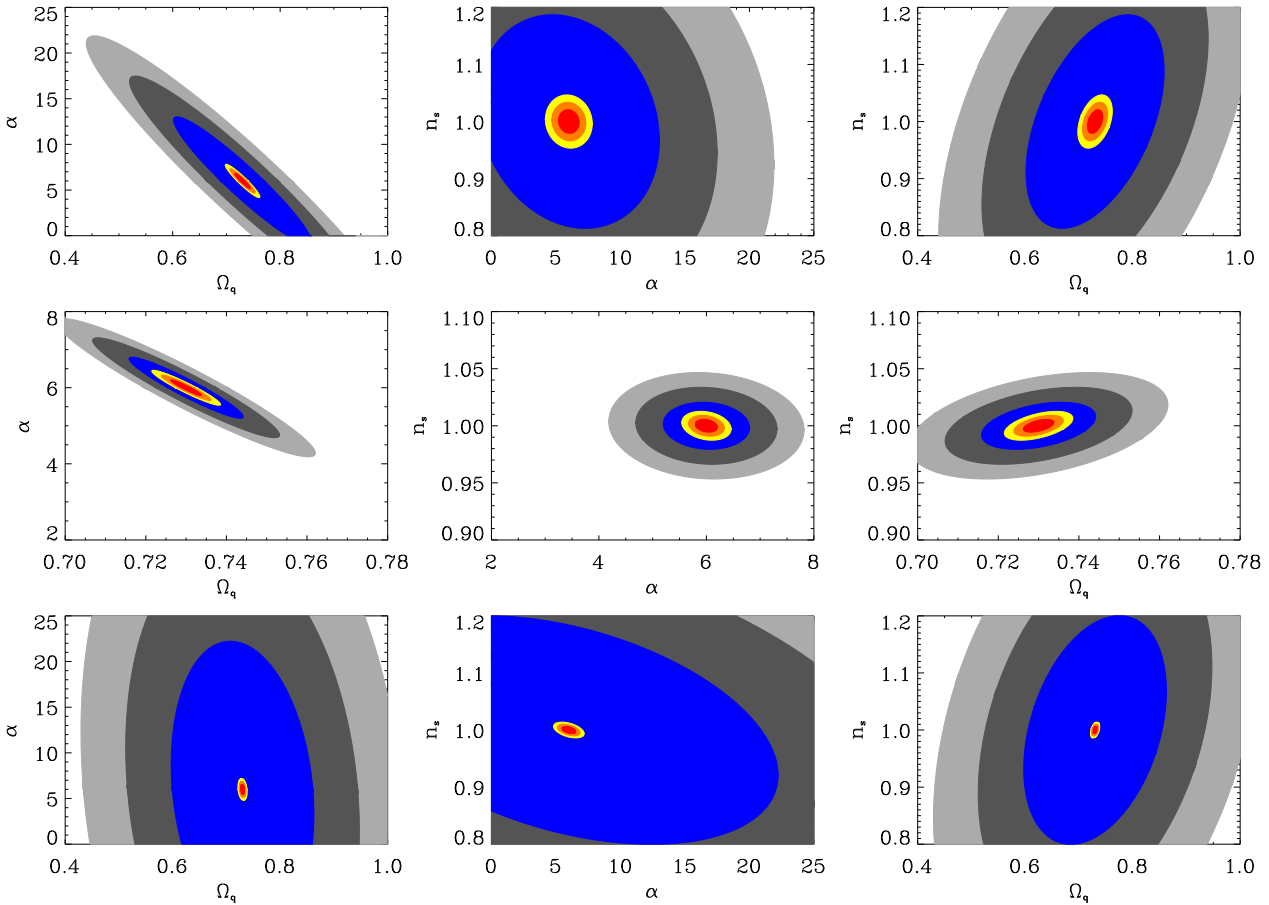}
 \caption{Fisher analysis of weak lensing (top-hat variance) for Ratra-Peebles
 models, considering a CFHTLS-wide like survey and two space-based surveys
 layout (68, 95\%, and 99\% C.L.). Upper panels, analogous to the the likelihood
 analysis of Sec.~\ref{sec3} which also assumes all the other parameters fixed,
 compare a CFHTLS-wide like survey (wider ellipses) with a deep space-based survey
 with $A=1,000$~deg$^2$, $n_\mathrm{gal}=50$~gal/arcmin$^2$, $\sigma_e=0.3$. Middle
 panels compare the deep space-based survey with a wider but shallow space-based survey
 with $A=20,000$~deg$^2$, $n_\mathrm{gal}=35$~gal/arcmin$^2$, $\sigma_e=0.3$.
 Bottom panels compare the CFHTLS-wide like survey with the type~I space-based
 survey, but marginalizing over $\tau_\mathrm{reion}$, left to vary. Compared
 with the first line, the last one shows the effect of a wider, more realistic
 parameters space. See \S~\ref{sec5} for discussion.}
\label{fig:5c}
\end{figure*}

As mentioned before, we may compute, a posteriori, the $\sigma_8$ value for each CMB normalized
model. The obtained $\sigma_8$ iso-contours in the $(\Omega_{\q 0},\alpha)$ plane are shown in
Fig.~\ref{fig:isol} (bottom panels), setting $n_s=1$. The $\sigma_8$ range that corresponds to
the confidence levels found in this work, is in agreement with current cosmic shear $\sigma_8$
constraints; see Hoekstra \etal (2005) for a recent result and van Waerbeke \& Mellier (2003)
for a compilation of results. This shows the normalizations on the CMB and at $z=0$ are compatible.
It is interesting to note that the directions of the $\Omega_{\q 0}-\alpha$ degeneracy for lensing
data are lines of constant $\sigma_8$. Being the curvature of the universe kept fixed, a strong
constraint in $\sigma_8$ implies a strong constraint in $\Omega_{\q 0}$, through the well-known
$\sigma_8-\Omega_{\mat 0}$ degeneracy, but only if $\Omega_{\q 0}$ and $\alpha$ are not much correlated.
This is what we observe in the SUGRA case. In general, a contour in the $(\sigma_8,\Omega_\mat)$
plane will move as a function of $\alpha$ and a strong constraint in $\sigma_8$ does not necessarily
imply a strong constraint in $\Omega_{\q 0}$. This is what is obtained for the Ratra-Peebles case.

The computational tools we have developed (see~\cite{cmbslow00}; \cite{SUR04}) allow
us to compute, in the same framework, distance modulus, CMB anisotropies and weak lensing -
cosmic shear effects. Hence, in principle there is no problem to combine Sn~Ia and lensing
data with CMB data. As already noticed, we have restricted here to a small parameter space.
Indeed, a joint analysis with CMB data require a wider parameter space, possibly using the
so-called ``normal parameters'' (see e.g. \cite{CMBfit}), an option left for a future study.
In such a case, we would integrate our pipeline with a Markov chain Monte Carlo code
(\cite{tereno04}) developed for the likelihood analysis of cosmic shear. Furthermore, we
will be able to include also the analysis of nucleosynthesis constraints, by means of a
suitable code for quintessence models (\cite{cocetal}) which is able to deal with ordinary and
extended quintessence models like the Boltzmann code and the lensing code used for this study.


\section{Weak lensing: prospects on future data}\label{sec5}

Beyond CFHTLS, several next generation cosmic shear surveys are proposed in order to pin
down $w$ and the dark energy properties with exquisite details. It is interesting to explore
the capabilities of such surveys to constrain the models discussed in this paper.

A comprehensive study of all projects and observing strategies is however beyond the scope of this
work, so we deliberately focus on rather simple concepts where both sky coverage and depth are
increased, assuming systematics related to shape measurement can be discarded. In particular, we
consider two possible layouts achievable by space-based missions. We will indicate by
\begin{itemize}
 \item{\it type~I}, a deep survey that would cover about 1,000~deg$^2$, one magnitude deeper than
the CFHTLS-wide providing an effective galaxy number density around 50~gal/arcmin$^2$;
 \item{\it type~II}, a wider but shallow survey, covering 20,000~deg$^2$ and yielding
35~gal/arcmin$^2$.
\end{itemize}
Furthermore, we assume an intrinsic ellipticity distribution of galaxies similar to the one
observed with current cosmic shear surveys with HST.

A Fisher analysis of Ratra-Peebles and SUGRA models, restricted to the parameters
$(\Omega_{\q 0},\alpha)$ keeping fixed the others, was performed around a fiducial model defined
by $(\alpha,\Omega_{\q 0},n_s,h,\tau_{\rm{reion}},\Omega_{\rm baryon0}h^2) =(6,0.73,1,0.72,0.17,
0.024)$. The results are presented in Fig.~\ref{fig:5b}. Weak lensing surveys (blue ellipses)
together with Sn~Ia ``gold'' set (green ellipses) and CMB WMAP-1yr data (red ellipses) show that
both quintessence parameters can in principle be determined with a 10\% accuracy. The degeneracy
with respect to Sn~Ia and CMB would be almost totally broken. Is is worth noticing that, since all
cosmological parameters but the quintessence ones are kept fixed, we have to take with care the
astonishing gain with respect to a CFHTLS-wide like survey (larger ellipses) achievable by a space
mission of type~II (smaller ellipses).

On scales larger than 10~degrees, the flat sky approximation used in \S~\ref{sec3a} does not hold
anymore. Indeed, an angular distance of 15~degrees on a sphere has a $1\%$ deviation from the same
distance on a plane, and for larger scales spherical harmonics must be considered
(\cite{stebbins}). These effects are not taken into account in the results of Fig.~\ref{fig:5b},
where the gain observed between the two weak lensing ellipses essentially corresponds to
Eq.~(\ref{like}), together with contributions from the fact that measurements from two surveys
observing at different scale ranges have different cosmic variances and different degrees of
freedom when fitting models to the data. In particular, a strategy of dividing the covered surveys
areas in patches of 100~deg$^2$ is assumed. Hence, given the large ratio between both sky coverages,
this is the dominant factor in the gain.

For indicative purposes, and allowing for the effect of systematic discussed in \S~\ref{sec4a}, we
restrict the Fisher analysis of weak lensing (top-hat variance) to Ratra-Peebles models,
considering a CFHTLS-wide like survey and both the space-based surveys of type~I and type~II; see
Fig.~\ref{fig:5c}. As in \S~\ref{sec4a}, we evaluate the parameters space $(\alpha,\Omega_{\q
0},n_s)$. In the upper line, analogous to the likelihood analysis of Sec.~\ref{sec3}, which also
assumes all the other cosmological parameters fixed, we compare the CFHTLS-wide like survey
(wider ellipses) with the space-based surveys of type~I (smaller ellipses). The middle line is
analogue, but comparing space-based surveys of type~I (wider ellipses) with the space-based
surveys of type~II (smaller ellipses); notice that the apparent rotation of the ellipses is simply
due to a rescaling of the axes. As for quintessence parameters, a space-based survey of type~II
gives 99\%~C.L. contours approximatively 3 times smaller than those achievable by a type~I survey.
Indeed more cosmological parameters have to be taken into account. As an example, allowing the
reionization optical depth $\tau_\mathrm{reion}$ to vary, the likelihood contours get strongly
modified; see the third line of Fig.~\ref{fig:5c}, where we compare a CFHTLS-wide like and
type~II space-based surveys, marginalizing over $\tau_\mathrm{reion}$. They get larger and the
degeneracies directions are changed. The strong impact of the reionization optical depth in
cosmic shear results comes from the degeneracy with between $\tau_\mathrm{reion}$ and the
normalization of the spectrum.

The great predictive power of space-based surveys will allow to simultaneously constrain a large
number of parameters with a good precision, even if not so high as the one showed in the analysis
of a small number of parameters. However, it is important to notice that we have not taken into
account the possible use of tomography based on a decomposition of lensing data into several
lensed/source planes, nor any use of higher order statistics than the top-hat shear variance.
So, even if several issues have been neglected in the present study, our conclusions are likely
not over-optimistic.


\section{Conclusions}\label{sec6}

In this article, we have investigated the constraints set by weak lensing, supernovae and CMB data
on two families of quintessence models.

In such a situation where a physical model is fully specified, we can treat both the background
and perturbation evolution without any ambiguity, in particular when dealing with high redshift
data. Such an approach is thus complementary, as discussed in Section~2, to those based on a
parameterization of the dark energy sector, in particular when trying to infer constraints on a
physical model from those on the parameters of the equation of state. This also enables to get rid
of the pivot redshift problem when combining different data sets. From a more theoretical point of
view, such models cannot lead to an equation of state $w<-1$, contrary to an arbitrary
parameterization. It is thus interesting to determine whether there is a tension in the data when
such a  physical constraint is imposed, as would be concluded from various studies indicating that
$w<-1$ is favored.

To achieve this task, we have used a set of numerical tools that allow to compute background, CMB
and lensing signatures of a large class of cosmological models including quintessence and some
extensions such as scalar-tensor theories. We have focused our analysis on three cosmological
parameters, the index of the primordial power spectrum and two parameters describing the quintessence
models, along with an extra-parameter for the sources distribution (see \S~\ref{sec3c}). Although
one can criticize such a small parameter space, it is sufficient to give an idea of the parameter
space available for quintessence models and to discuss how weak lensing data can improve the
constraint on dark energy. This choice was also driven by numerical limitations but our analysis
will be extended to a larger set of parameters in a near future.

We have normalized the initial power spectrum to the CMB so that $\sigma_8$ is now a prediction of
the models and is not used for normalization. Note also that we do not require to specify an
analytical form for the transfer function. Weak lensing predictions are also sensitive to the
linear to non-linear mapping and we have discussed the effect of such mapping on the constraints
with care. In particular, we have shown that, while the parameters of the primordial spectrum are
sensitive, those of the dark energy sector remain robust. We have also tested the possibility to
cut the weak lensing data sets (such as the CFHTLS-wide) in order to reduce the influence of the
non-linear regime.

This analysis is the first one using CFHTLS data to study the dark energy and illustrates the
complementarity of these observations with other data sets. To finish, we have also forecast how
space-based wide field imagers will improve our knowledge of dark energy. In particular, we have
considered two possible strategies, the first deeper and the second wider but shallower. The
latter turns out to be more suited to track dark energy as far as cosmic shear is concerned. The
constraints on the two classes of quintessence models considered in this article are shown on
Fig.~\ref{fig:4a} and Table~\ref{tab:4}. They can be summarized as follows: For a flat universe and
a quintessence inverse power law potential with slope $\alpha$, we get $\alpha<1$ and
$\Omega_\mathrm{Q0}=0.75^{+0.03}_{-0.04}$ at 95\% confidence level, whereas $\alpha=2^{+18}_{-2}$,
$\Omega_{\q0}=0.74^{+0.03}_{-0.05}$ when including supergravity corrections.

In the future, we plan to improve this analysis by first comparing it to a similar analysis based
on a parameterization of the equation of state, by enlarging the parameter space, by addressing
more carefully the problem of the redshift distribution of galaxies and by shifting from a grid
method to an MCMC method.

\begin{acknowledgements}
We thank Nabila Aghanim, Karim Benabed, Francis Bernardeau, Daniel Eisenstein, Bernard Fort,
Simon Prunet, Alexandre R\'efr\'egier and Jim Rich. CS acknowledges ``Fondazione Angelo Della
Riccia'' and ``Fondazione Ing. Aldo Gini'' for financial support. IT is partly funded by the
CNRS/ANR research grant "ECOSSTAT", contract number  ANR-05-BLAN-0283-04 .LF thanks the "European
Association for Research in Astronomy" training site (EARA) and the European Community for
the Marie Curie doctoral fellowship MEST-CT-2004-504604. LVW, HH are supported by the Natural
Sciences and Engineering Research Council (NSERC), the Canadian Institute for Advanced Research
(CIAR) and the Canadian Foundation for Innovation (CFI). We thank the CNRS-INSU and the French
Programme National de Cosmologie for their support to the CFHTLS cosmic shear program.

\end{acknowledgements}


\end{document}